\documentclass[oribibl,envcountsame]{llncs}

\usepackage{latexsym}
\usepackage{amstext}
\usepackage{enumerate}

\title{The Complexity of Boolean Constraint Isomorphism\thanks{%
Research supported in part by grants
NSF-INT-9815095/\protect\linebreak[0]DAAD-315-PPP-g\"u-ab,
NSF-CCR-0311021, DAAD D/0205776, DFG VO 630/5-1, and by an RIT FEAD grant.
}}

\author{Elmar B\"ohler\inst{1}
\and Edith Hemaspaandra\inst{2}
\and Stef{}fen Reith\inst{3}
\and Heribert Vollmer\inst{4}}

\institute{Theoretische Informatik,
Universit\"{a}t W\"{u}rzburg,
Am Hubland,
D-97074 W\"{u}rzburg, Germany,
e-mail: boehler@informatik.uni-wuerzburg.de
\and
Department of Computer Science,
Rochester Institute of Technology,
Rochester, NY 14623, U.S.A.,
e-mail: eh@cs.rit.edu
\and
LengfelderStr. 35b,
D-97078 W\"{u}rzburg, Germany,
e-mail: streit@streit.cc
\and
Theoretische Informatik, Universit\"at Hannover, Appelstr.\ 4, 
D-30167 Hannover, Germany,
email: vollmer@thi.uni-hannover.de
}

\newtheorem{fact}[theorem]{Fact}
\newtheorem{ourclaim}[theorem]{Claim}
\newtheorem{observation}[theorem]{Observation}
\renewenvironment{proof}{\noindent {\bf Proof.}\quad}{\qed}
\newenvironment{prooftext}[1]{\noindent {\bf Proof of #1.}\quad}{\qed}

\newcommand{\CSP}{\mbox{\rm CSP}}

\newcommand{\XOR}{\mbox{\rm XOR}}

\newcommand{\oneinthree}{\mbox{\rm OneInThree}}
\newcommand{\majority}{\mbox{\it majority}}
\newcommand{\nf}{\mbox{\it nf}}
\newcommand{\OR}{\mbox{\rm OR}}

\newcommand{\ISO}{\mbox{\rm ISO}}

\newcommand{\GI}{\mbox{\rm GI}}

\newcommand{\condition}{\ |\ }
\newcommand{\littlep}{{p}}
\newcommand{\manyone}{\ensuremath{\,\leq_{m}^{{\littlep}}\,}}

\newcommand{\p}{{\rm P}}

\newcommand{\xor}{\oplus}

\newcommand{\cnto}[1]{\ensuremath{\#_{1}(#1)}}
\newcommand{\eqd}{\ensuremath{=_{\mathrm{\scriptscriptstyle def}}}}

\newcount\penum
\def\pitem{\ifnum\penum>1\par\fi\textbf{\the\penum.} \advance\penum by1}

\newenvironment{penumerate}
  {\penum=1}{}

\begin{document}

\maketitle

\begin{abstract}
We consider the Boolean constraint isomorphism problem, that is,
the problem of determining whether two sets of Boolean constraint
applications can be made equivalent by renaming the variables.
We show that depending on the set of allowed constraints, the problem is either
coNP-hard and GI-hard, equivalent to graph isomorphism, or polynomial-time
solvable. This establishes a complete classification of the complexity
of the problem, and moreover, it identifies exactly all those cases in
which Boolean constraint isomorphism is polynomial-time many-one
equivalent to graph isomorphism, the best-known and best-examined
isomorphism problem in theoretical computer science.
\end{abstract}

\section{Introduction}

Constraint satisfaction problems (or, constraint networks) were
introduced in 1974 by U.~Montanari to solve computational problems
related to picture processing \cite{mon74}.  It turned out that they
form a broad class of algorithmic problems that arise naturally in
different areas \cite{kol03}.  Today, they are ubiquitous in computer
science (database query processing, circuit design, network
optimization, planning and scheduling, programming languages),
artificial intelligence (belief maintenance and knowledge based
systems, machine vision, natural language understanding), and
computational linguistics (formal syntax and semantics of natural
languages).

A constraint satisfaction instance is given by a set of variables, a
set of values that the variables may take (the so-called
\emph{universe}), and a set of constraints. A constraint restricts the
possible assignments of values to variables; formally a $k$-place
constraint is a $k$-ary relation over the universe. The most basic
question one is interested in is to determine if there is an
assignment of values to the variables such that all constraints are
satisfied.

This problem has been studied intensively in the past decade from a
computational complexity point of view. In a particular case, that of
2-element universes, a remarkable complete classification was
obtained, in fact already much earlier, by Thomas Schaefer
\cite{Sch78}. Note that in this case of a Boolean universe, the
variables are propositional variables and the constraints are Boolean
relations. A constraint satisfaction instance, thus, is a
propositional formula in conjunctive normal form where, instead of the
usual clauses, arbitrary Boolean relations may be used. In other
words, the constraint satisfaction problem here is the satisfiability
problem for generalized propositional formulas.  Obviously the
complexity of this problem depends on the set $\cal C$ of constraints
allowed, and is therefore denoted by $\CSP({\cal C})$ ($\cal C$ will
always be finite in this paper). In this way we obtain an infinite
family of NP-problems, and Schaefer showed that each of them is either
NP-complete or polynomial-time solvable. This result is surprising,
since by Ladner's Theorem \cite{lad75b} there is an infinite number of
complexity degrees between P and NP (assuming P $\neq$ NP), and
consequently it is well conceivable that the members of an infinite
family of problems may be located anywhere in this hierarchy. Schaefer
showed that for the generalized satisfiability problem this is not the
case: Each $\CSP({\cal C})$ is either NP-complete, that is in the
highest degree, or in the lowest degree P. Therefore his result is
called a \emph{dichotomy theorem}.

For larger universes, much less is known. Satisfiability of constraint
networks is always in NP, and for large families of allowed sets of
constraints, NP-completeness was proven while for others, tractability
(i.e., polynomial-time algorithms) was obtained. Research in this
direction was strongly influenced by the seminal papers
\cite{jecogy97,feva98}, and many deep and beautiful results have been
proven since then, see, e.\,g.,
\cite{jecoco98,KoVa98,jecogy99,bujekr00,bukrje01,bobujekr03}. 
Only recently, a dichotomy theorem for the complexity of
satisfiability of constraint networks over 3-element universes was
published \cite{bul02b}, but for larger domains such a complete
classification still seems to be out of reach. For Boolean universes,
however, a number of further computational problems have been
addressed and in most cases, dichotomy theorems were obtained. These
problems concern, among others, the problems to count how many
satisfying solutions an instance has \cite{CrHe96}, to enumerate all
satisfying solutions \cite{CrHe97}, to determine in certain ways
optimal satisfying assignments \cite{Cr95,KiKo01,revo03}, to determine
if there is a unique satisfying assignment \cite{Ju99}, learnability
questions related to propositional formulas \cite{dal00}, the inverse
satisfiability problem \cite{KaSi98}, and the complexity of
propositional circumscription \cite{kiko01c,duhe03}.  Results about
approximability of optimization problems related to Boolean CSPs
appeared in \cite{Zw98,KSTW97}.  We point the reader to the monograph
\cite{CrKhSu00} that discusses much of what is known about Boolean
constraints.

In this paper, we address a problem that is not a variation of
satisfiability, namely, the \emph{isomorphism problem} for Boolean
constraints. Perhaps the most prominent isomorphism problem in
computational complexity theory is the \emph{graph isomorphism
problem}, GI, asking given two graphs if they are isomorphic. Graph
isomorphism has been well studied because it is one of the very few
problems in NP neither known to be NP-complete nor known to be in P
(in fact, there is strong evidence that GI is not NP-complete, see
\cite{KoScTo93}); thus GI may be in one of the ``intermediate
degrees'' mentioned above.

Another isomorphism problem studied intensively in the past few years
is the propositional formula isomorphism. This problem asks, given two
propositional formulas, if there is a renaming of the variables that
makes both equivalent. The history of this problem goes back to the
19th century, where Jevons and Clifford, two mathematicians, were
concerned with the task to construct formulas or circuits for all
$n$-ary Boolean functions, but since there are too many ($2^{2^n}$) of
them they wanted to identify a small set of Boolean circuits from
which all others could then be obtained by some simple
transformation. This problem has been referred to since as the
``Jevons-Clifford-Problem.'' One of the transformations they used was
renaming of variables (producing an \emph{isomorphic} circuit),
another one was first renaming the variables and then negating some of
them (producing what has been called a \emph{congruent}
circuit). Hence it is important to know how many equivalence classes
for isomorphism and congruence there are, and how to determine if two
circuits or formulas are isomorphic or congruent.  (A more detailed
discussion of these developments can be found in
\cite[pp.~6--8]{thi00}.) However, the exact complexity of the
isomorphism problem for Boolean circuits and formulas (the congruence
problem turns out to be of the same computational complexity;
technically: both problems are polynomial-time many-one equivalent) is still
unknown: It is trivially hard for the class coNP (of all complements
of NP-problems) and in $\Sigma_2^p$ (the second level of the
polynomial hierarchy), and Agrawal and Thierauf showed that it is most
likely not $\Sigma_2^p$-hard (that is, unless the polynomial hierarchy
collapses, an event considered very unlikely by most
complexity-theorists) \cite{AgTh00}.

In this paper we study the Boolean formula isomorphism problem
restricted to formulas in the Schaefer sense, in other words: the
\emph{isomorphism problem for Boolean constraints}. In a precursor,
the present authors showed that this problem is either coNP-hard (the
hard case, the same as for general formula isomorphism) or reducible
to the graph isomorphism problem (the easy case)
\cite{BoHeReVo02}. This result is not satisfactory, since it leaves
the most interesting questions open: Are there ``really easy'' cases
for which the isomorphism problem is tractable (that is, in P)? What
exactly are these? And are the remaining cases which reduce to graph
isomorphism actually equivalent to GI?

The present paper answers these questions affirmatively. To state
precisely our main result (Theorem~\ref{t:isotrich}) already here
(formal definitions of the relevant classes of constraints will be
given in the next section), constraint isomorphism is coNP-hard and
GI-hard for classes $\cal C$ of constraints that are neither Horn nor
anti-Horn nor affine nor bijunctive, it is in in P if $\cal C$ is both
affine and bijunctive, and in all other cases, the isomorphism problem
is \emph{equivalent to graph isomorphism}. This classification holds
for constraint applications with as well as without constants. As in
the case of Schaefer's dichotomy, we thus obtain simple criteria to
determine, given $\cal C$, which of the three cases holds.  This
theorem gives a complete classification of the computational
complexity of Boolean constraint isomorphism. Moreover, it determines
exactly all those cases of the Boolean constraint isomorphism problem
that are equivalent to graph isomorphism, the most prominent and
probably most studied isomorphism problem so far.

The next section formally introduces constraint satisfaction problems
and the relevant properties of constraints. Section~\ref{s:tricho} then
contains the proof of our main theorem: In
Section~\ref{s:iso-upper} we identify those classes of constraints for
which isomorphism is in P and Section~\ref{s:gihard} contains the
main technical contribution of this paper proving GI-hardness for all
other cases.

\section{Preliminaries}

We start by formally introducing constraint problems.
The following section is
essentially from~\cite{BoHeReVo02}, following the standard notation
developed in~\cite{CrKhSu00}.

\begin{definition}\rm
\begin{enumerate}
\item A {\em constraint} $C$ (of arity $k$) is a Boolean function from
$\{0,1\}^k$ to $\{0,1\}$.
\item If $C$ is a constraint of arity $k$, and $x_1, x_2, \dots, x_k$
are (not necessarily distinct) variables, then $C(x_1, x_2, \dots,
x_k)$ is a {\em constraint application of $C$}.  In this paper, we
view a constraint application as a Boolean function on a specific set
of variables.  Thus, for example, $x_1 \vee x_2 = x_2 \vee x_1$
\item If $C$ is a constraint of arity $k$, and for $1 \leq i \leq k$,
$x_i$ is a variable or a constant (0 or 1), then $C(x_1, x_2, \dots,
x_k)$ is a constraint application of $C$ {\em with constants}.
\item If $A$ is a constraint application [with constants], and $X$ a
set of variables that includes all variables that occur in $A$, we say
that $A$ is a constraint application [with constants] {\em over
variables $X$}.  Note that we do not require that every element of $X$
occurs in $A$.
\end{enumerate}
\end{definition}

The complexity of Boolean constraint problems depends on those properties of
constraints that we define next.

\begin{definition}\rm
Let $C$ be a constraint.
\begin{itemize}
\item $C$ is {\em 0-valid} if $C(\vec{0}) = 1$.  Similarly, $C$ is
{\em 1-valid} if $C(\vec{1}) = 1$.
\item $C$ is {\em Horn} (or {\em weakly negative})
[\emph{anti-Horn} (or {\em weakly positive})] 
if $C$ is equivalent to a CNF
formula where each clause has at most one positive [negative]
literal.
\item $C$ is {\em bijunctive} if $C$ is equivalent to a 2CNF formula.
\item $C$ is {\em affine} if $C$ is equivalent to an XOR-CNF formula.
\item $C$ is {\em 2-affine} (or, affine with width 2) if $C$ is
equivalent to a XOR-CNF formula such that every clause contains at
most two literals.
\end{itemize}
Let $\cal C$ be a finite set of constraints.  We say ${\cal C}$ is
0-valid, 1-valid, Horn, anti-Horn, bijunctive, or affine
if {\em every} constraint $C\in{\cal C}$ is 0-valid,
1-valid, Horn, anti-Horn, bijunctive, or affine,
respectively. Finally, we say that ${\cal C}$ is {\em Schaefer} if
${\cal C}$ is Horn or anti-Horn or affine or bijunctive.
\end{definition}

The question studied in this paper is that of whether a set of
constraint applications can be made equivalent to a second set of
constraint applications using a suitable renaming of its variables. We
need some definitions.

\begin{definition}\rm
\begin{enumerate}
\item Let $S$ be a set of constraint applications with constants over
variables $X$ and let $\pi$ be a permutation of $X$.  By
\emph{$\pi(S)$} we denote the set of constraint applications that
results when we replace simultaneously all variables $x$ in $S$ by
$\pi(x)$.
\item Let $S$ be a set of constraint applications over variables $X$.
The number of satisfying assignments of $S$, $\cnto{S}$, is
defined as $||\{\,I
\condition I $ is an assignment to all variables in $X$ that satisfies
every constraint application in $S\,\}||$.
\end{enumerate}
\end{definition}

The isomorphism problem for Boolean constraints, first defined and examined in
\cite{BoHeReVo02} is formally defined as follows.

\begin{definition}\rm
\begin{enumerate}
\item $\ISO({\cal C})$ is the problem of, given two sets $S$ and $U$
of constraint applications of ${\cal C}$ over variables $X$, to decide
whether $S$ and $U$ are isomorphic, i.e., whether there exists a
permutation $\pi$ of $X$ such that $\pi(S)$ is equivalent to $U$.
\item $\ISO_c({\cal C})$ is the problem of, given two sets $S$ and $U$
of constraint applications of ${\cal C}$ with constants over variables
$X$, to decide whether $S$ and $U$ are isomorphic.
\end{enumerate}
\end{definition}

B\"ohler et al.\ obtained results about the complexity of the just-defined
problem that, interestingly, pointed out relations to another
isomorphism problem: the graph isomorphism problem (GI).

\begin{definition}\rm
\GI\ is the problem of, given two graphs $G$ and $H$, to determine
whether $G$ and $H$ are isomorphic, i.e., whether there exists a
bijection $\pi\colon V(G) \rightarrow V(H)$ such that for all $v,w \in
V(G)$, $\{v, w\} \in E(G)$ iff $\{\pi(v), \pi(w)\} \in E(H)$.  Our
graphs are undirected, and do not contain self-loops.  We also assume
a standard enumeration of the edges, and will write $E(G) = \{e_1,
\ldots, e_m\}$.
\end{definition}

GI is a problem that is in NP, not known to be in P, and not
NP-complete unless the polynomial hierarchy collapses.  For details,
see, for example,~\cite{KoScTo93}.  Recently, Tor{\'a}n showed that GI
is hard for NL, PL, Mod$_k$L, and DET under logspace many-one
reductions~\cite{To00}.  Arvind and Kurur showed that GI is in the
class SPP~\cite{ArKu02}, and thus, for example in $\oplus P$.

The main result from \cite{BoHeReVo02} can now be stated as follows.

\begin{theorem}\label{isodichth}
Let ${\cal C}$ be a set of constraints.  If ${\cal C}$ is Schaefer,
then $\ISO({\cal C})$ and $\ISO_c({\cal C})$ are polynomial-time
many-one reducible to \GI, otherwise, $\ISO({\cal C})$ and
$\ISO_c({\cal C})$ are {\rm coNP}-hard.
\end{theorem}

Note that if $\cal C$ is Schaefer the isomorphism problems $\ISO({\cal
C})$ and $\ISO_c({\cal C})$ cannot be coNP-hard, unless NP = coNP.
(This follows from Theorem~\ref{isodichth} and the fact that GI is in
NP.)  Under the (reasonable) assumption that NP $\neq$ coNP, and that
GI is neither in P, nor NP-complete, Theorem~\ref{isodichth} thus
distinguishes a hard case (coNP-hard) and an easier case (many-one
reducible to GI).

B\"ohler et al.{} also pointed out that there are some bijunctive,
Horn, or affine constraint sets ${\cal C}$ for which actually
$\ISO({\cal C})$ and $\ISO_c({\cal C})$ are equivalent to graph
isomorphism. On the other hand, certainly there are ${\cal C}$ for
which $\ISO({\cal C})$ and $\ISO_c({\cal C})$ are in $\p$. In the upcoming
section we will completely classify the complexity of $\ISO({\cal C})$
and $\ISO_c({\cal C})$, obtaining for which ${\cal C}$ exactly we are
equivalent to \GI\, and for which ${\cal C}$ we are in $\p$.

\section{A Classification of Boolean Constraint Isomorphism}
\label{s:tricho}

The main result of the present paper is a complete
complexity-theoretic classification of the isomorphism problem for
Boolean constraints.

\begin{theorem}\label{t:isotrich}
Let ${\cal C}$ be a finite set of constraints.
\begin{enumerate}
\item If ${\cal C}$ is not Schaefer, then $\ISO({\cal C})$
and $\ISO_c({\cal C})$ are {\rm coNP}-hard and \GI-hard.
\item If ${\cal C}$ is Schaefer and not 2-affine, then $\ISO({\cal C})$
and $\ISO_c({\cal C})$ are polynomial-time many-one equivalent to \GI.
\item Otherwise, ${\cal C}$ is 2-affine and $\ISO({\cal C})$ and
$\ISO_c({\cal C})$ are in {\rm P}.
\end{enumerate}
\end{theorem}

The rest of this section is devoted to a proof of this theorem and organized as follows.  The coNP
lower-bound part from Theorem~\ref{t:isotrich} follows from
Theorem~\ref{isodichth}.  In Section~\ref{s:iso-upper} we will prove
the polynomial-time upper bound if ${\cal C}$ is 2-affine
(Theorem~\ref{th:isop}). The \GI\ upper bound if ${\cal C}$ is
Schaefer again is part of Theorem~\ref{isodichth}.  In
Section~\ref{s:gihard} we will show that $\ISO_c({\cal C})$ is GI-hard
if ${\cal C}$ is not 2-affine (Theorems~\ref{th:gitonotaffine}
and~\ref{th:gitonotbijunctive}).
Theorem~\ref{t:noconst} finally shows that $\ISO({\cal C})$ is GI-hard
if ${\cal C}$ is not 2-affine.

\subsection{Upper Bounds}\label{s:iso-upper}

A central step in our way of obtaining upper bounds is to bring sets
of constraint applications into a unique normal form. This approach is
also followed in the proof of the ${\rm coIP}[2]^{\rm NP}$ upper 
bound\footnote{Here ${\rm IP}[2]$ means an interactive proof system where 
there are two messages exchanged between the verifier and the prover.}
for the isomorphism problem for Boolean formulas~\cite{AgTh00}
and
the $\GI$ upper bound from Theorem~\ref{isodichth} \cite{BoHeReVo02}.

\begin{definition}\rm
\label{d:nf}
Let ${\cal C}$ be a set of constraints. $\nf$ is a {\em normal form
function for ${\cal C}$} if and only if for all sets $S$ and $U$ of
constraint applications of ${\cal C}$ with constants over variables
$X$, and for all permutations $\pi$ of $X$,
\begin{enumerate}
\item $\nf(S,X)$ is a set of Boolean functions over variables $X$,
\item $S \equiv \nf(S,X)$ (here we view $S$ as a set of
Boolean functions, and define equivalence for such sets as logical
equivalence of corresponding propositional formulas),
\item $\nf(\pi(S),X) = \pi(\nf(S,X))$, and
\item if $S \equiv U$, then $\nf(S,X) = \nf(U,X)$ (here, ``$=$'' is
equality between sets of Boolean functions).
\end{enumerate}
It is important to note that $\nf(S,X)$ is not necessarily
a set of constraint applications of ${\cal C}$ with constants.
\end{definition}

An easy property of the definition is that  $S \equiv U$ iff
$\nf(S,X) = \nf(U,X)$.  Also, it is not too hard to observe that using
normal forms removes the need to check whether two sets of constraint
applications with constants are equivalent, more precisely: $S$ is
isomorphic to $U$ iff there exists a permutation $\pi$ of $X$ such
that $\pi(\nf(S)) = \nf(U)$.

There are different possibilities for normal forms.  The one used by
\cite{BoHeReVo02} is the maximal equivalent set of constraint
applications with constants, defined by $\nf(S,X)$ to be the set of
all constraint applications $A$ of ${\cal C}$ with constants over
variables $X$ such that $S \rightarrow A$.  For the P upper bound for
2-affine constraints, we use the normal form described in the following
lemma, whose proof can be found in Appendix~\ref{proof:l:affinenf}. 
Note that this normal form is not necessarily a set of 2-affine
constraint applications with constants.

\begin{lemma}
\label{l:affinenf}
Let ${\cal C}$ be a set of 2-affine constraints.
There exists a polynomial-time computable
normal form function $\nf$ for ${\cal C}$ such that
for all sets $S$ of
constraint applications of ${\cal C}$ with
constants over variables $X$, the following hold:
\begin{enumerate}
\item If $S \equiv 0$, then $\nf(S,X) = \{0\}$.
\item If $S \not \equiv 0$, then 
$\nf(S,X) =  \{\overline{Z}, {O}\} \cup
\bigcup_{i = 1}^\ell  \{ (X_i \wedge \overline{Y_i})
\vee (\overline{X_i} \wedge Y_i) \}$,
where $Z, O,  X_1, Y_1, \ldots, X_\ell,Y_\ell$ are pairwise
disjoint subsets of $X$ such that
$X_i \cup Y_i \neq \emptyset$ for all $1 \leq i \leq \ell$,
and for $W$ a set of variables, $W$ in a formula denotes
$\bigwedge W$, and $\overline{W}$ denotes $\neg \bigvee W$.
\end{enumerate}
\end{lemma}

Making use of the normal form, it is not too hard to prove our claimed
upper bound.

\begin{theorem}
\label{th:isop}
Let ${\cal C}$ be a set of constraints.  If ${\cal C}$ is 2-affine,
then $\ISO({\cal C})$ and $\ISO_c({\cal C})$ are in {\rm P}.
\end{theorem}

\begin{proof}
Let $S$ and $U$ be two sets of constraint applications of ${\cal C}$
and let $X$ be the set of variables that occur in $S \cup U$.  Use
Lemma~\ref{l:affinenf} to bring $S$ and $U$ into normal form.  Using
the first point in that lemma, it is easy to check whether $S$ or $U$
are equivalent to $0$.  For the remainder of the proof, we now suppose
that neither $S$ nor $U$ is equivalent to $0$.  Let $Z, O, X_1, Y_1,
\ldots, X_\ell,Y_\ell$ and $Z', {O}', {X}_1', {Y}_1',\ldots, {X}_k',
{Y}_k'$ be subsets of $X$ such that:
\begin{enumerate}
\item $Z, O, X_1, Y_1, \ldots, X_\ell,Y_\ell$ are pairwise disjoint
and ${Z}', {O}', {X}_1', {Y}_1', \ldots, {X}_k', {Y}_k'$ are pairwise
disjoint,
\item $X_i \cup Y_i \neq \emptyset$ for all $1 \leq i \leq \ell$ and
${X}_i' \cup {Y}_i' \neq \emptyset$ for all $1 \leq i \leq k$,
\item $\nf(S,X) = \{\overline{Z}, {O}\} \cup \bigcup_{i = 1}^\ell \{
(X_i \wedge \overline{Y_i}) \vee (\overline{X_i} \wedge Y_i) \}$, and
$\nf(U,X) = \{\overline{{Z}'}, {O}'\} \cup \bigcup_{i = 1}^k \{ (
{X}_i' \wedge \overline{{Y}_i'}) \vee (\overline{{X}_i'} \wedge
{Y}_i') \}$.
\end{enumerate}

We need to determine whether $S$ is isomorphic to $U$.  Since $\nf$ is
a normal form function for ${\cal C}$, it suffices to check if there
exists a permutation $\pi$ on $X$ such that $\pi(\nf(S,X)) =
\nf(U,X)$. Note that
$$\pi(\nf(S,X))  =  \{\overline{\pi(Z)}, \pi(O)\} \cup
\bigcup_{i = 1}^\ell  \{ (\pi(X_i) \wedge \overline{\pi(Y_i)})
\vee (\overline{\pi(X_i)} \wedge \pi(Y_i)) \}.$$
It is immediate that $\pi(\nf(S,X)) = \nf(U,X)$ if and only if
\begin{itemize}
\item $\ell = k$, $\pi(Z) = {Z}'$, $\pi(O) = {O}'$, and
\item
$\{\{\pi(X_1), \pi(Y_1)\},\dots,\{\pi(X_\ell),\pi(Y_\ell)\}\} 
=\{\{{X}_1', {Y}_1'\}, \dots, \{{X}_\ell',{Y}_\ell'\}\}$.
\end{itemize}

Since $Z, O, X_1, Y_1, \ldots, X_\ell,Y_\ell$ are pairwise disjoint
subsets of $X$, and since ${Z}', {O}', {X}_1', {Y}_1', \ldots,
{X}_k',{Y}_k'$ are pairwise disjoint subsets of $X$, it is easy to see
that there exists a permutation $\pi$ on $X$ such that $\nf(\pi(S),X)
= \nf(U,X)$ if and only if
\begin{itemize}
\item
$\ell = k$, $||Z|| = ||{Z}'||$, $||O|| = ||{O}'||$, and
\item
$[\{||X_1||, ||Y_1||\}, \dots, \{||X_\ell||,||Y_\ell||\}] =
[\{||{X}_1'||, ||{Y}_1'||\}, \dots, \{||{X}_\ell'||,||{Y}_\ell'||\}]$;\\
here $[\cdots]$ denotes a multi-set.
\end{itemize}

It is easy to see that the above conditions can be verified
in polynomial time. It follows that
$\ISO({\cal C})$ and $\ISO_c({\cal C})$ are in P.
\end{proof}

\subsection{GI-hardness}\label{s:gihard} 

In this section, we will prove that if ${\cal C}$ is not 2-affine,
then \GI\ is polynomial-time many-one reducible to $\ISO_c({\cal C})$
and $\ISO({\cal C})$.
As in the upper bound proofs of the previous section, we will often
look at certain normal forms.  In this section, it is often convenient
to avoid constraint applications that allow duplicates.

\begin{definition}\rm
Let ${\cal C}$ be a set of constraints.
\begin{enumerate}
\item $A$ is a constraint application of ${\cal C}$
{\em without duplicates} if there exists a constraint $C \in {\cal C}$ of
arity $k$ such that $A = C(x_1, \ldots, x_k)$, where $x_i \neq x_j$ for
all $i \neq j$.
\item
Let $S$ be a set of constraint applications of ${\cal C}$
[without duplicates] over variables $X$.
We say that $S$ is a maximal set of constraint applications
of ${\cal C}$ [without duplicates]
over variables $X$ if for all constraint applications $A$
of ${\cal C}$ [without duplicates] over variables $X$,
if $S \rightarrow A$, then $A \in S$.

If $X$ is the set of variables occurring in $S$, we will
say that $S$ is a maximal set of constraint applications
of ${\cal C}$ [without duplicates].
\end{enumerate}
\end{definition}

The following lemma is easy to see.

\begin{lemma}
\label{l:max}
Let ${\cal C}$ be a set of constraints.
Let $S$ and $U$ be maximal sets of constraint applications of ${\cal C}$
over variables $X$ [without duplicates].
Then $S$ is isomorphic to $U$ iff there exists a permutation $\pi$ of
$X$ such that $\pi(S) = U$.
\end{lemma}

Note that if ${\cal C}$ is not 2-affine, then 
${\cal C}$ is not affine, or ${\cal C}$ is affine and not bijunctive.
We will first look at some very simple non-affine constraints.

\begin{definition}[{\cite[p.\ 20]{CrKhSu00}}]\rm\leavevmode
\begin{enumerate}
\item $\OR_0$ is the constraint $\lambda xy. x \vee y$.
\item $\OR_1$ is the constraint $\lambda xy. \overline{x} \vee y$.
\item $\OR_2$ is the constraint $\lambda xy. \overline{x} \vee \overline{y}$.
\item $\oneinthree$ is the constraint $\lambda xyz .  (x \wedge
\overline{y} \wedge \overline{z}) \vee (\overline{x} \wedge y \wedge
\overline{z}) \vee (\overline{x} \wedge \overline{y} \wedge z)$.
\end{enumerate}
\end{definition}

As a first step in the general GI-hardness proof, we show that GI
reduces to some particular constraints. The reduction of $\GI$ to
$\ISO(\{\OR_0\})$ already appeared in \cite{BRS95}. Reductions in the
other cases follow similar patterns. The proofs can be found in
Appendix~\ref{proof:l:gitorotten}.

\begin{lemma}
\label{l:gitoor}\label{l:gitoorxor}\label{l:gitorotten}
\begin{enumerate}
\item\label{l:1}
\GI\ is polynomial-time many-one reducible to $\ISO(\{\OR_i\})$,
$i\in\{0,1,2\}$.
\item\label{l:2}
Let $h$ be the 4-ary constraint $h(x,y,x',y') = (x \vee y) \wedge (x
\oplus x') \wedge (y \oplus y')$.  \GI\ is polynomial-time many-one
reducible to $\ISO(\{h\})$.
\item\label{l:3}
Let $h$ be a 6-ary constraint $h(x,y,z,x',y',z') = \oneinthree(x,y,z)
\wedge (x \oplus x') \wedge (y \oplus y') \wedge (z \oplus z')$.  Then
\GI\ is polynomial-time many-one reducible to $\ISO(\{h\})$.
\end{enumerate}
\end{lemma}

The constraints $\OR_0$, $\OR_1$, and $\OR_2$ are the simplest
non-affine constraints.  However, it is not enough to show that GI
reduces to the isomorphism problem for these simple cases.  In order
to prove that GI reduces to the isomorphism problem for all sets of
constraints that are not affine, we need to show that all such sets
can ``encode'' a finite number of simple cases.

Different encodings are used in the lower bound proofs for different
constraint problems. All encodings used in the literature
however, allow the introduction of auxiliary variables.
In~\cite{CrKhSu00}, Lemma 5.30, it is shown that if $C$ is not affine,
then $C$ plus constants can encode $\OR_0$, $\OR_1$, or $\OR_2$. This implies that,
for certain problems, lower bounds for $\OR_0$, $\OR_1$, or $\OR_2$
transfer to $C$ plus constants.  However, their encoding uses auxiliary variables,
which means that lower bounds for the isomorphism problem don't
automatically transfer. For sets of constraints that are not affine,
we will be able to use part of the proof of~\cite{CrKhSu00}, Lemma
5.30, but we will have to handle auxiliary variables explicitly, which
makes the constructions much more complicated.

\begin{theorem}
\label{th:gitonotaffine}
If ${\cal C}$ is not affine, then \GI\ is polynomial-time many-one
reducible to $\ISO_c({\cal C})$.
\end{theorem}

\begin{proof}
First suppose that ${\cal C}$ is weakly negative and weakly positive.
Then ${\cal C}$ is bijunctive \cite{CrHe96}.  From the proof
of~\cite[Lemma 5.30]{CrKhSu00}
it follows that there exists a constraint
application $A(x,y,z)$ of ${\cal C}$ with constants such that
$A(0,0,0) = A(0,1,1) = A(1,0,1) = 1$ and $A(1, 1, 0) = 0$.  Since
${\cal C}$ is weakly positive, we also have that $A(1, 1, 1) =
1$. Since ${\cal C}$ is bijunctive, we have that $A(0,0,1) = 1$.  The
following truth-table summarizes all possibilities (this is a
simplified version of~\cite{CrKhSu00}, Claim 5.31).
\begin{center}
\begin{tabular}{|c|c|c|c|c|c|c|c|c|c|c|}
$xyz$&000&001&010&011&100&101&110&111\\
\hline
$A(x,y,z)$&1&$1$&$a$&1&$b$&1&0&1
\end{tabular}
\end{center}

Thus we obtain $A(x,x,y) = (\overline{x} \vee y)$, and the
result follows from Lemma~\ref{l:gitoor}.\ref{l:1}.

So, suppose that ${\cal C}$ is not weakly negative or not weakly
positive.  We follow the proof of~\cite{CrKhSu00}, Lemma 5.30.  From
the proof of~\cite{CrKhSu00}, Lemma 5.26, it follows that there exists
a constraint application $A$ of ${\cal C}$ with constants such that
$A(x,y) = \OR_0(x,y)$, $A(x,y) = \OR_2(x,y)$, or $A(x,y) = x \xor y$.
In the first two cases, the result follows from
Lemma~\ref{l:gitoor}.\ref{l:1}.

Consider the last case. From the proof of~\cite{CrKhSu00}, Lemma 5.30,
there exist a set $S(x,y,z,x',y',z')$ of ${\cal C}$ constraint
applications with constants and a ternary function $h$ such that
$S(x,y,z,x',y',z') = h(x,y,z) \wedge (x \oplus x') \wedge (y \oplus
y') \wedge (z \oplus z')$, $h(000) = h(011) = h(101) = 1$, and $h(110)
= 0$.

The following truth-table summarizes all possibilities:
\begin{center}
\begin{tabular}{|c|c|c|c|c|c|c|c|c|c|c|}
$xyz$&000&001&010&011&100&101&110&111\\
\hline
$h(x,y,z)$&1&$a$&$b$&1&$c$&1&0&$d$
\end{tabular}
\end{center}

We will first show that in most cases, there exists a set $U$ of
constraint applications of ${\cal C}$ with constants such that
$U(x,y,x',y') = (x \vee y) \wedge (x \oplus x') \wedge (y \oplus y')$.
In all these cases, the result follows from
Lemma~\ref{l:gitoorxor}.\ref{l:2} above.

\begin{itemize}
\item $b = 0, d = 1$.  In this case, $S(x,y,x,x',y',x') =
(x \vee y') \wedge (x \oplus x') \wedge (y \oplus y')
 = (x \vee y') \wedge (x \oplus x') \wedge (y \oplus y')$.
\item $b = 1, d = 0$.  In this case, $S(x,y,x,x',y',x') =
(x' \vee y') \wedge (x \oplus x') \wedge (y \oplus y')$.
\item $c = 0, d = 1$.  In this case, $S(x,y,y,x',y',y') =
(x' \vee y) \wedge (x \oplus x') \wedge (y \oplus y')$.
\item $c = 1, d = 0$.  In this case, $S(x,y,y,x',y',y') =
(x' \vee y') \wedge (x \oplus x') \wedge (y \oplus y')$.
\item $b = c = 1$.  In this case, $S(x,y,0,x',y',1) =
(x' \vee y') \wedge (x \oplus x') \wedge (y \oplus y')$.
\item $b = c = d = 0; a = 1$.  In this case, $S(0,y,z,1,y',z') =
(y' \vee z) \wedge (y \oplus y') \wedge (z \oplus z')$.
\end{itemize}
The previous cases are analogous to the cases from the proof of
\cite{CrKhSu00}, Claim 5.31.  However, we have to explicitly add the
$\oplus$ conjuncts to simulate the negated variables used there, which
makes Lemma~\ref{l:gitoorxor}.\ref{l:2} necessary.

The last remaining case is the case where $a = b = c = d = 0$.  In the
proof of~\cite{CrKhSu00}, Claim 5.31, it suffices to note that
$(\overline{y} \vee z) = \exists! x h(x,y,z)$. But, since we are
looking at isomorphism, we cannot ignore auxiliary variables.  Our
result uses a different argument and follows from Lemma~\ref{l:gitorotten}.3 above and the
observation that $S(x,y,z,x',y',z') = \oneinthree(x,y,z') \wedge (x
\oplus x') \wedge (y \oplus y') \wedge (z \oplus z')$.
\end{proof}

In the case where ${\cal C}$ is affine but not 2-affine, we first show
GI-hardness of a particular constraint and then turn to the general
result. 
(The proofs, using similar constructions as in the proofs of
Lemma~\ref{l:gitorotten} and Theorem~\ref{th:gitonotaffine}, is
given in  Appendix~\ref{proof:l:gitothreeaffine} and
\ref{proof:th:gitonotbijunctive}.)

\begin{lemma}
\label{l:gitothreeaffine}
Let $h$ be the 6-ary constraint such that $h(x,y,z,x',y',z') = (x
\oplus y \oplus z) \wedge (x \oplus x') \wedge (y \oplus y') \wedge (z
\oplus z')$.  \GI\ is polynomial-time many-one reducible to
$\ISO(\{h\})$.
\end{lemma}

\begin{theorem}
\label{th:gitonotbijunctive}
If ${\cal C}$ is affine and not bijunctive, then
\GI\ is polynomial-time many-one reducible to $\ISO_c({\cal C})$.
\end{theorem}

Finally, to finish the proof of statement 2 of
Theorem~\ref{t:isotrich}, it remains to show GI-hardness of
$\ISO({\cal C})$ for ${\cal C}$ not 2-affine.
In Appendix~\ref{s:rem-const} we show that it is
possible to remove the introduction of constants in the previous
constructions of this section.

\medskip

\noindent{\bf Acknowledgements:}
We would like to thank Lane Hemaspaandra for helpful conversations
and suggestions, and the anonymous referees for helpful comments.

\goodbreak

\newpage

\appendix

\section{Proof of Lemma \ref{l:affinenf}}
\label{proof:l:affinenf}

Let $S$ be a set of constraint applications of ${\cal C}$ with
constants over variables $X$.  Since ${\cal C}$ is 2-affine, we can in
polynomial time check whether $S \equiv 0$.  If so, $\nf(S,X) =
\{0\}$.  Now suppose that $S$ is not equivalent to $0$.  Let ${\cal
D}$ be the set of all unary and binary 2-affine constraints, i.e.,
${\cal D} = \{\lambda a. a, \lambda a. \overline{a}, \lambda ab. a
\oplus b, \lambda ab. \neg(a \oplus b)\}$.

Let $S''$ be the set of all constraint applications $A$ of ${\cal D}$
with constants over variables $X$ such that $S \rightarrow A$. In
other words, we are using the maximal equivalent set normal form used
in \cite{BoHeReVo02}, that we described in the paragraph preceding
Lemma~\ref{l:affinenf}.  It follows from \cite{BoHeReVo02} that $S''$
is computable in polynomial time. Certainly, $S \equiv S''$, since
every constraint application in $S$ can be written as the conjunction
of constraint applications of ${\cal D}$. We will now show how to compute
$\nf(S,X)$.

Let $Z$ be the set of those variables in $X$ that, when set to $1$,
make $S''$ equivalent to $0$. Let $O$ be the set of those variables in $X$
that, when set to $0$, make $S''$ equivalent to $0$.  Note that $S''
\rightarrow \overline{Z} \wedge O$.  Let $S'$ be the set of constraint
applications that is obtained from $S''$ by setting all elements of
$Z$ to $0$ and all elements of $O$ to 1.  Note that $S'' \equiv
\overline{Z} \wedge O \wedge S'$.

Let $\widehat{S}$ be the set of all constraint applications from $S'$
that do not contain constants.  We claim that $\widehat{S} \equiv S'$.
For suppose that $\alpha$ is an assignment that satisfies
$\widehat{S}$, but $\alpha$ does not satisfy $S'$.  Then there is a
constraint application with constants $A \in S' \setminus
\widehat{S}$ such that $\alpha$ does not satisfy $A$.  Note that $A$
must contain a constant, since $A \not \in \widehat{S}$.  $A$ must
contain a variable, since $A$ is satisfiable (since $S$ is
satisfiable).  But then $A$ contains exactly one occurrence of exactly
one variable, and thus $A$ is equivalent to $x$ or $\overline{x}$ for some
variable $x$. But then $x$ would have been put in $Z$ or $O$, and $x$
would not occur in $S'$.

So, $S \equiv \overline{Z} \wedge O \wedge \widehat{S}$, and every
element of $\widehat{S}$ is of the form $x \oplus y$ where $x,y \in X$
and $x \neq y$ or of the form $\neg (x \oplus y)$, where $x, y \in X$ and
$x \neq y$ or of the form $\neg (x \oplus x)$, where $x \in X$.
Note that it is possible that $\widehat{S} = \emptyset$ and that
all variables in $X$ occur in $S$.

Also note that, since $S''$ contains all of its implicates that are of
the right form, for every three distinct variables $x$, $y$, and $z$,
\begin{itemize}
\item if $(x \oplus y) \in \widehat{S}$ and 
$(y \oplus z) \in \widehat{S}$, then  $\neg(x \oplus z) \in \widehat{S}$,
\item if $\neg(x \oplus y) \in \widehat{S}$ and 
$(y \oplus z) \in \widehat{S}$, then  $(x \oplus z) \in \widehat{S}$, and
\item if $\neg(x \oplus y) \in \widehat{S}$ and 
$\neg(y \oplus z) \in \widehat{S}$, then  $\neg (x \oplus z)
\in \widehat{S}$.
\end{itemize}
So, $\widehat{S}$ is closed under a form of transitivity.

Partition $\widehat{S}$ into $S_1, \ldots, S_{\ell}$, where $S_1,
\ldots, S_{\ell}$ are minimal sets that are pairwise disjoint with
respect to occurring variables.  Since the $S_i$s are minimal, and not
equivalent to 0, it follows from the observation above about the
closure of $\widehat{S}$, that for every pair of distinct variables
$x,y$ in $S_i$, exactly one of $(x \oplus y)$ and $\neg (x \oplus y)$
is in $S_i$.  For every $i$, $1 \leq i \leq \ell$, let $x_i$ be an
arbitrary variable that occurs in $S_i$.  Let $X_i = \{y \ | \
\neg(x_i \oplus y) \in S_i\}$ and let $Y_i = \{y \ | \ x_i \oplus y
\in S_i\}$. Then $X_i \cap Y_i = \emptyset$ and $X_i \cup Y_i$ = the
variables that occur in $S_i$.

It is easy to see that $S_i \equiv \{ (X_i \wedge \overline{Y_i})
\vee (\overline{X_i} \wedge Y_i) \}$.

We claim that
\[\nf(S,X)  =  \{\overline {Z},  {O}\} \cup
\bigcup_{i = 1}^\ell  \{ (X_i \wedge \overline{Y_i})
\vee (\overline{X_i} \wedge Y_i) \} \]
fulfills the criteria of Lemma~\ref{l:affinenf}.

First of all, it is clear that $\nf$ is computable in polynomial time.
{From} the observations above,
it follows that $Z, O, X_1, Y_1, \ldots, X_\ell,Y_\ell$ are
pairwise disjoint subsets of $X$ such that
\begin{enumerate}
\item $X = Z \cup O \cup \bigcup_{i = 1}^{\ell}(X_i \cup Y_i)$, and
\item $X_i \cup Y_i \neq \emptyset$ for all $1 \leq i \leq \ell$,
\end{enumerate}
that $\nf(S,X) \equiv S$,
and that $\nf(\pi(S),X) = \pi(\nf(S),X)$, for all
permutations $\pi$ of $X$.

It remains to show that if $U$ is a set of constraint applications of
${\cal C}$ with constants, and $S \equiv U$, then $\nf(S,X) =
\nf(U,X)$.

Let ${Z}', {O}', {X}_1', {Y}_1',\ldots, {X}_k',{Y}_k'$ be subsets of
$X$ such that:
\begin{enumerate}
\item ${Z}', {O}', {X}_1', {Y}_1',\ldots, {X}_k',{Y}_k'$ are pairwise
disjoint,
\item $X = {Z}' \cup {O}' \cup \bigcup_{i = 1}^{k}({X}_i' \cup
{Y}_i')$,
\item ${X}_i' \cup {Y}_i' \neq \emptyset$ for all $1 \leq i \leq
\ell$,
\item
$\nf(U,X) = \{\overline{Z'}, {O}'\} \cup \bigcup_{i = 1}^k \{ ({X}_i'
\wedge \overline{{Y}_i'}) \vee (\overline{{X}_i'} \wedge {Y}_i') \}$.
\end{enumerate}

Since $S \equiv \nf(S,X)$, $U \equiv \nf(U,X)$, and $S \equiv U$, it
follows that $\nf(S,X) \equiv \nf(U,X)$.  From this, it is immediate
that $Z = {Z}'$ and that $O = {O}'$.  In addition, for any two
variables $x,y \in X$:
\begin{itemize}
\item
($\{x,y\} \in X_i$ or $\{x,y\} \in Y_i$ for some $i$) iff ($S$ is
satisfiable iff $x \equiv y$) iff ($\{x,y\} \in {X}_j'$ or $\{x,y\}
\in {Y}_j'$ for some $j$), and
\item
($\{x,y\} \cap X_i \neq \emptyset$ and $\{x, y\} \cap Y_i \neq
\emptyset$ for some $i$) iff ($S$ is satisfiable iff $x \not \equiv
y$) iff ($\{x,y\} \cap {X}_j' \neq \emptyset$ and $\{x, y\} \cap
{Y}_j' \neq \emptyset$ for some $j$).
\end{itemize}
This implies that $\nf(U,X) = \nf(S,X)$, which completes the proof.

\section{Graph Isomorphism for Restricted Graphs}
\label{proof:l:girestriction}

When reducing from GI, it is often useful to assume that the graphs
have certain properties.  The following lemma (whose proof is given in
Appendix~\ref{proof:l:girestriction}) shows that the complexity of GI
for certain restricted classes of graphs does not decrease.

\begin{lemma}
\label{l:girestriction}
\GI\ is polynomial-time many-one reducible to the graph isomorphism
problem for pairs of graphs $G$ and $H$ such that for some
$n$, $G$ and $H$ have the same set of vertices $\{1, \ldots, n\}$,
$G$ and $H$ have the same number of edges,
every vertex in $G$ and $H$ has degree at least two
(i.e.,  is incident with at least two edges),
and $G$ and $H$ do not contain triangles.
\end{lemma}

\begin{proof}
Let $G$ and $H$ be graphs.  If $G$ and $H$ do not contain
the same number of vertices, or if $G$ and $H$ do not
contain the same number of edges,
or if $G$ and $H$ do not contain the same number of
isolated vertices, then $G$ is not isomorphic to $H$.

So suppose that $G$ and $H$ have the same number of vertices,
the same number of edges, and the same number of isolated
vertices. Let $G_1$ be the graph that results if we remove
all isolated vertices from $G$.  Let $H_1$ be the graph that
results if we remove all isolated vertices from $H$.  Then
$G_1$ and $H_1$ have the same number of vertices and the same
number of edges, all vertices in $G_1$ and $H_1$ have degree at least
one, and $G$ is isomorphic to $H$ iff $G_1$ is isomorphic to $H_1$.
Without loss of generality, assume that $G_1$ has at least 3 vertices.

We will now ensure that no vertex has degree less than two.
Let $v_0$ be a new vertex.  Define $G_2$ as follows:
$V({G}_2) = V({G}_1) \cup \{v_0\}$,
$E({G}_2) = E({G}_1) \cup 
\{\{v_0, v\} \condition v \in V({G}_1)\}$. 
Define $H_2$ in the same way, i.e.,
$V({H}_2) = V({H}_1) \cup \{v_0\}$,
$E({H}_2) = E({H}_1) \cup 
\{\{v_0, v\} \condition v \in V({H}_1)\}$. 
Note that $G_2$ and $H_2$ have the same number of vertices and the
same number of edges, and that all vertices in $G_2$ and $H_2$ have degree
at least two. In addition, it is easy to see that
$G_1$ is isomorphic to $H_1$ iff $G_2$ is isomorphic to $H_2$:
If $\pi$ is an isomorphism from $G_2$ to $H_2$, then we can define
an isomorphism $\rho$ from $G_1$ to $H_1$ as follows:
For all $v  \in V(G_1)$, $\rho(v) = \pi(v)$ if $\pi(v) \neq v_0$,
and $\rho(v) = \pi(v_0)$ if $\pi(v) = v_0$.

Next, we will remove triangles.
Define ${G}_3$ as
follows: $V({G}_3) = V({G}_2) \cup E({G}_2)$,
$E({G}_3) = \{\{v,w\} \condition v \in V({G}_2),
w \in E({G}_2), v \in w\}$.
Define $H_3$ in the same way, i.e., 
$V({H}_3) = V({H}_2) \cup E({H}_2)$,
$E({H}_3) = \{\{v,w\} \condition v \in V({H}_2),
w \in E({H}_2), v \in w\}$.
Note that $G_3$ and $H_3$ are triangle-free graphs
with the same number of vertices and the same number of edges,
and that all vertices in $G_3$ and $H_3$ have degree
at least two.  We claim that
$G_2$ is isomorphic to $H_2$ iff $G_3$ is isomorphic to $H_3$.
The left-to-right direction is immediate.  The right-to-left direction
follows since an isomorphism from $G_3$ to $H_3$ maps
$V(G_2)$ to $V(H_2)$, since these are exactly the 
vertices at even distance from a vertex of degree greater than 2.
(Here we use the fact that the degree of $v_0$ is greater than 2.)

Let $n$ be the number of vertices of $G_3$ and $H_3$. Rename
the vertices in $G_3$ and $H_3$ to $\{1, 2, \ldots, n\}$. This proves
Lemma~\ref{l:girestriction}.
\end{proof}

\section{Proof of Lemma~\ref{l:gitorotten}}
\label{proof:l:gitorotten}

\begin{penumerate}
\pitem As mentioned, the polynomial-time many-one reduction of $\GI$
to $\ISO(\{\OR_0\})$ was already published in \cite{BRS95}.  We will
first recall this proof, since the proofs of the other cases will be
similar though more complicated.  Let $\widehat{G}$ be a graph and let
$V(\widehat{G}) = \{1, 2, \ldots , n\}$.  We encode $\widehat{G}$ in
the obvious way as a set of constraint applications $S(\widehat{G}) =
\{x_i \vee x_j \condition \{i,j\} \in E(\widehat{G})\}$.  It is easy
to see that $S(\widehat{G})$ is a maximal set of constraint
applications of $\OR_0$.  If $G$ and $H$ are two graphs without
isolated vertices and with vertex set $\{1, 2, \ldots, n\}$, then $G$
is isomorphic to $H$ if and only if there exists a permutation $\pi$
of $\{x_1, \ldots, x_n\}$ such that $\pi(S(G)) = S(H)$.  By
Lemma~\ref{l:max} it follows that $G$ is isomorphic to $H$ if and only
if $S(G)$ is isomorphic to $S(H)$.

If we negate all occurring variables in $S(\widehat{G})$, i.e.,
$S(\widehat{G}) = \{\overline{x_i} \vee \overline{x_j} \condition
\{i,j\} \in E(\widehat{G})\}$, we obtain a reduction from $\GI$ to
$\ISO(\{\OR_2\})$.

It remains to show that $\GI$ is reducible to $\ISO(\{\OR_1\})$.  Note
that the obvious encoding $\{x_i \vee \overline{x_j} \condition
\{i,j\} \in E(\widehat{G})\}$ does not work, since a 3-vertex graph
with edges $\{1,2\}$ and $\{1,3\}$ will be indistinguishable from a
3-vertex graph with edges $\{1,2\}$, $\{1,3\}$, and $\{2,3\}$.

We solve this problem by using a slightly more complicated encoding.
Let $E(\widehat{G}) = \{e_1, \ldots, e_m\}$.  Define $S(\widehat{G}) =
\{x_i \vee \overline{y_k}, x_j \vee \overline{y_k} \condition e_k =
\{i,j\}\}$.  We claim that if $G$ and $H$ are two graphs without
isolated vertices with vertex set $\{1, 2, \ldots, n\}$, then $G$ is
isomorphic to $H$ if and only if $S(G)$ is isomorphic to $S(H)$.

The left-to-right direction is immediate.  For the converse, note that
for all $\widehat{G}$, $S(\widehat{G})$ is a maximal set of constraint
applications of $\OR_1$.  Thus, by Lemma~\ref{l:max}, if $S(G)$ is
isomorphic to $S(H)$, there exists a permutation $\pi$ of the
variables such that $\pi(S(G)) = S(H)$. Since the $x_i$ variables are
exactly those variables that occur positively in $S(G)$ and $S(H)$,
$\pi$ maps $x$-variables to $x$-variables, and thus induces an
isomorphism from $G$ to $H$.

\pitem
We use a similar encoding as in the first case above.  Let
$\widehat{G}$ be a graph and let $V(\widehat{G}) = \{1, 2, \ldots ,
n\}$. We encode $\widehat{G}$ as the following set of constraint
applications of $h$: $S(\widehat{G}) = \{h(x_i,x_j,x_i',x_j')
\condition \{i,j\} \in E(\widehat{G})\}$.

Let $X = \{x_1, \ldots, x_n, x_1', \ldots, x_n'\}$.  Note that for all
variables $x,y \in X$, ($S(\widehat{G}) \rightarrow (x \vee y)$ and
$S(\widehat{G}) \not \rightarrow (x \oplus y)$) iff there exists an
edge $\{i,j\} \in E$ such that $\{x,y\} = \{x_i,x_j\}$.

Let $G$ and $H$ be two graphs without isolated vertices and with
vertex set $\{1, 2, \ldots, n\}$.  We claim that $G$ is isomorphic to
$H$ if and only if $S(G)$ is isomorphic to $S(H)$. The left-to-right
direction is immediate, since an isomorphism $\pi$ from $G$ to $H$
induces an isomorphism $\rho$ from $S(G)$ to $S(H)$, by letting
$\rho(x_i) = x_{\pi(i)}$ and $\rho(x_i') = x_{\pi(i)}'$.

For the converse, suppose that $\rho$ is a permutation of $X$ such
that $\rho(S(G)) \equiv S(H)$.  As noted above, for all $\{i,j\} \in
E(G)$, it holds that $S(H) \rightarrow (\rho(x_i) \vee \rho(x_j))$ and
$S(H) \not \rightarrow (\rho(x_i) \oplus \rho(x_j))$.  Again by the
observation above, there exists an edge $\{k, \ell\} \in E(H)$ such
that $\{\rho(x_i),\rho(x_j)\} = \{x_k,x_\ell\}$. Thus, $\rho$ maps
$x$-variables to $x$-variables.  Let $\pi(i) = j$ iff $\rho(x_i) =
x_j$.  It is easy to see that $\pi$ is an isomorphism from $G$ to $H$.

\pitem 
Let $\widehat{G}$ be a graph such that $V(\widehat{G}) = \{1, 2,
\ldots, n\}$ and $E(\widehat{G}) = \{e_1, \ldots, e_m\}$.  Define
$S(\widehat{G})$ as follows: $S(\widehat{G}) =
\{h(x_i,x_j,y_k,x_i',x_j',y_k') \condition e_k = \{i,j\}\} \cup \{x_i
\oplus x_i' \condition 1 \leq i \leq n\} \cup \{y_i \oplus y_i'
\condition 1 \leq i \leq m\}.$

Define $U(\widehat{G})$ as follows: $U(\widehat{G}) =
\{\oneinthree(x_i,x_j,y_k) \condition e_k = \{i,j\}\} \cup \{x_i
\oplus x_i' \condition 1 \leq i \leq n\} \cup \{y_i \oplus y_i'
\condition 1 \leq i \leq m\}.$ Clearly, $U(\widehat{G})$ is equivalent
to $S(\widehat{G})$.  Let $X$ be the set of all variables that occur
in $U(\widehat{G})$.

We will use the following claim whose proof will be given in
Appendix~\ref{proof:cl:rottennf}.

\begin{ourclaim}
\label{cl:rottennf}
The set of all constraint applications of
$\oneinthree$ without duplicates that occur
in $U(\widehat{G})$ is a maximal set of
constraint applications of $\oneinthree$
over variables $X$ without duplicates, where $X$ is the set of
all variables that occur in $U(\widehat{G})$.
\end{ourclaim}

Let $G$ and $H$ be graphs without isolated vertices, with
vertex set $\{1, 2, \ldots, n\}$,  
and with the same number of edges. Also assume that every
vertex in $G$ and $H$ is incident with at least two edges
(see Lemma~\ref{l:girestriction}).
Let $E(G) = \{e_1, \ldots, e_m\}$ and let
$E(H) = \{e'_1, \ldots, e'_m\}$.

We claim that $G$ is isomorphic to $H$ iff $S(G)$ is isomorphic to
$S(H)$.  It suffices to show that $G$ is isomorphic to $H$ iff $U(G)$
is isomorphic to $U(H)$.

The left-to-right direction is trivial, since an isomorphism between
the graphs induces an isomorphism between sets of constraint
applications as follows.  If $\pi: V \rightarrow V$ is an isomorphism
from $G$ to $H$, then we can define an isomorphism $\rho$ from $U(G)$
to $U(H)$ as follows:
\begin{itemize}
\item
$\rho(x_i) = x_{\pi(i)}$, $\rho(x'_i) = x'_{\pi(i)}$, for $i \in V$.
\item
For $e_k = \{i,j\}$, $\rho(y_k) = y_{\ell}$, and $\rho(y'_k) =
y'_{\ell}$, where $e'_{\ell} = \{\pi(i),\pi(j)\}$.
\end{itemize}

For the converse, suppose that $\rho$ is an isomorphism from $U(G)$ to
$U(H)$.  Let $U'(G)$ and $U'(H)$ be the sets of all constraint
applications of $\oneinthree$ without duplicates that occur in $U(G)$
and $U(H)$, respectively.

{From} Claim~\ref{cl:rottennf} and Lemma~\ref{l:max}, it follows that
$\rho(U'(G)) = U'(H)$.  Note that the $x$-variables in $U'(G)$ and
$U'(H)$ are exactly those variables that occur in at least two
constraint applications of $\oneinthree$ without constants.  Thus,
$\rho$ maps $x$-variables to $x$-variables. Likewise, $\rho$ maps
$y$-variables to $y$-variables. Let $\pi$ be the bijection on $\{1, 2,
\ldots, n\}$ defined by $\pi(i) = j$ iff $\rho(x_i) = x_j$.  We claim
that $\pi$ is an isomorphism from $G$ to $H$. First let $e_k =
\{i,j\}$. Thus, $\oneinthree(x_i, x_j,y_k) \in U'(G)$.  Then,
$\oneinthree(\rho(x_i), \rho(x_j), \rho(y_k)) \in U'(H)$.  Thus,
$\oneinthree(x_{\pi(i)}, x_{\pi(j)}, \rho(y_k)) \in U'(H)$.  This
implies that $\{\pi(i),\allowbreak \pi(j)\}$ is an edge in $H$.  For the
converse, suppose that $e'_k = \{\pi(i), \pi(j)\}$ is an edge in $H$.
Then $\oneinthree(x_{\pi(i)}, x_{\pi(j)}, y_k) \in U'(H)$.  Thus we conclude
$\oneinthree(\rho(x_i), \rho(x_j), y_k) \in U'(H)$ and hence
$\oneinthree(x_i, x_j, \rho^{-1}(y_k)) \in U'(G)$.  It follows that
$\{i,j\}$ is an edge in $G$.
\end{penumerate}

\section{Proof of Lemma~\ref{l:gitothreeaffine}}
\label{proof:l:gitothreeaffine}

Following \cite[p.\ 20]{CrKhSu00}, we use $\XOR_2$ to denote the
constraint $\lambda xy. x \oplus y$, and $\XOR_3$ to denote $\lambda
xyz. x \oplus y \oplus z$.

Let $\widehat{G}$ be a graph such that $V(\widehat{G}) = \{1, 2,
\ldots , n\}$, and $E(\widehat{G}) = \{e_1, e_2, \ldots, e_m\}$.  We
will use a similar encoding as in the proof of
Lemma~\ref{l:gitorotten}.  Again, propositional variable $x_i$ will
correspond to vertex $i$ and propositional variable $y_i$ will
correspond to edge $e_i$.

Define $S(\widehat{G})$ as follows: $S(\widehat{G}) =
\{h(x_i,x_j,y_k,x_i',x_j',y_k') \condition e_k = \{i,j\}\} \cup \{x_i
\oplus x_i' \condition 1 \leq i \leq n\} \cup \{y_i \oplus y_i'
\condition 1 \leq i \leq m\}.$

Define $U(\widehat{G})$ as follows: $U(\widehat{G}) = \{(x_i \oplus
x_j \oplus y_k), (x_i' \oplus x_j' \oplus y_k), (x'_i \oplus x_j
\oplus y'_k), (x_i \oplus x_j' \oplus y_k') \condition e_k = \{i,j\}\}
\cup \{x_i \oplus x_i' \condition 1 \leq i \leq n\} \cup \{y_i \oplus
y_i' \condition 1 \leq i \leq m\}.$ Clearly, $U(\widehat{G})$ is
equivalent to $S(\widehat{G})$.  Let $X$ be the set of variables
occurring in $U(\widehat{G})$.

The proof relies on the following claim, which shows that
$U(\widehat{G})$ is a maximal set of constraint applications of
$\{\XOR_2, \XOR_3\}$ without duplicates (a proof can be found in
Appendix~\ref{proof:cl:maxxor}):

\begin{ourclaim}
\label{cl:maxxor}
Let $\widehat{G}$ be a triangle-free graph such that
$V(\widehat{G}) = \{1,2, \ldots, n\}$,
$E(\widehat{G}) = \{e_1,e_2, \ldots, e_m\}$, and
every vertex has degree at least two.
Then $U(\widehat{G})$ is a maximal set
of constraint applications of $\{\XOR_2, \XOR_3\}$ without duplicates.
\end{ourclaim}

Let $G$ and $H$ be graphs such that
$V(G) = V(H) = \{1, \ldots, n\}$,
$E(G) = \{e_1, \ldots, e_m\}$,
$E(H) = \{e'_1, \ldots, e'_m\}$,
all vertices in $G$ and $H$ have degree at least two,
and $G$ and $H$ do not contain triangles.

We will show that $G$ is isomorphic to $H$ if and only if $S(G)$ is
isomorphic to $S(H)$.  It suffices to show that $G$ is isomorphic to
$H$ if and only if $U(G)$ is isomorphic to $U(H)$.

The left-to-right direction is trivial, since an isomorphism between
the graphs induces an isomorphism between sets of constraint
applications as follows.  If $\pi: V \rightarrow V$ is an isomorphism
from $G$ to $H$, then we can define an isomorphism $\rho$ from $U(G)$
to $U(H)$ as follows:
\begin{itemize}
\item
$\rho(x_i) = x_{\pi(i)}$,
$\rho(x'_i) = x'_{\pi(i)}$, for $i \in V$.
\item
For $e_k = \{i,j\}$,
$\rho(y_k) = y_{\ell}$, and
$\rho(y'_k) = y'_{\ell}$, where $e'_{\ell} = \{\pi(i),\pi(j)\}$.
\end{itemize}

For the converse, suppose that $\rho$ is an isomorphism from $U(G)$ to
$U(H)$.  By Lemma~\ref{l:max} and Claim~\ref{cl:maxxor}, $\rho(U(G)) =
U(H)$.  Note that every $y_i$ and $y'_i$ variable occurs in exactly
two constraint applications of $\XOR_3$ in $U(G)$ and $U(H)$, while
every $x_i$ and $x'_i$ variable occurs in at least four constraint
applications of $\XOR_3$ in $U(G)$ and $U(H)$ (since every vertex in
$G$ and $H$ is incident with at least two edges). From the $\XOR_2$
constraint applications, it is also immediate that for all $a \in
\{x_1, \ldots, x_n, y_1, \ldots, y_m\}$, there exists a $b \in \{x_1,
\ldots, x_n, y_1, \ldots, y_m\}$ such that $\{\rho(a), \rho(a')\} =
\{b, b'\}$.

Define $\pi$ as follows: $\pi(i) = j$ if and only if $\{\rho(x_i),
\rho(x_i')\} = \{x_j, x'_j\}$.  By the observations above, $\pi$ is
total and 1-1.  It remains to show that $\{i,j\} \in E(G)$ iff
$\{\pi(i),\pi(j)\} \in E(H)$.

Let $e_k = \{i,j\}$. Then $x_i \xor x_j \xor e_k \in U(G)$.
Thus, $\rho(x_i) \xor \rho(x_j) \xor \rho(y_k) \in U(H)$.
That is, $a \oplus b \oplus \rho(y_k) \in U(H)$ for
some $a \in \{x_{\pi(i)}, x'_{\pi(i)}\}$ and
$b \in \{x_{\pi(j)}, x'_{\pi(j)}\}$.
But that implies that $\rho(y_k) \in \{y_{\ell}, y'_{\ell}\}$ where 
$e'_{\ell} = \{\pi(i),\pi(j)\}$. This implies that
$\{\pi(i),\pi(j)\} \in E(H)$. For the converse, suppose that
$\{\pi(i),\pi(j)\} \in E(H)$. Then
$x_{\pi(i)} \xor x_{\pi(j)} \xor y_{\ell} \in U(H)$ for
$e'_{\ell} = \{\pi(i),\pi(j)\}$.  It follows that
$a \oplus b \oplus \rho^{-1}(y_{\ell}) \in U(G)$ for
some $a \in \{x_i, x'_i\}$ and
$b \in \{x_j, x'_j\}$.  By the form
of $U(G)$, it follows that $\{i,j\} \in E(G)$.

\section{Proof of Theorem~\ref{th:gitonotbijunctive}}
\label{proof:th:gitonotbijunctive}

Schaefer \cite{Sch78} characterized classes of Boolean constraints in
terms of closure properties. Important for us will be his characterization
of bijunctive constraints.

\begin{lemma}[\cite{Sch78}]\rm
\label{l:bfchar}
Let $f$ be a Boolean function of arity $k$.
$f$ is bijunctive if and only if
for all assignments $s, t, u \in \{0,1\}^k$ that satisfy $f$,
$\majority(s,t,u)$ (the vector obtained from $s,t,u$ by bitwise
majority) satisfies $f$.
\end{lemma}

\begin{proof} (of Theorem~\ref{th:gitonotbijunctive})

Recall from the proof of Theorem~\ref{th:gitonotaffine} that if
${\cal C}$ is not bijunctive, then ${\cal C}$ is not weakly positive
or not weakly negative.  As in the proof of that theorem, it follows
from the the proof of~\cite{CrKhSu00}, Lemma 5.26 that there exists a
constraint application $A$ of ${\cal C}$ with constants such that
$A(x,y) = \OR_0(x,y)$, $A(x,y) = \OR_2(x,y)$, or $A(x,y) = x \xor y$.
Since $A$ is affine, and $\OR_0$ and $\OR_2$ are not affine,
the first two cases cannot occur.

Consider the last case.  Let $B \in {\cal C}$ be a constraint that is
not bijunctive.  Let $k$ be the arity of $B$.  Following Schaefer's
characterization of bijunctive functions (see Lemma~\ref{l:bfchar}),
there exist assignments $s ,t, u \in \{0,1\}^k$ such that $B(s) = B(t)
= B(u) = 1$ and $B(\majority(s,t,u)) = 0$.  In addition, since ${\cal
C}$ is affine, using Schaefer's characterization of affine functions,
$B(s \oplus t \oplus u) = 1$.

Let $\widehat{B}(x,y,z,x',y',z') = B(x_{1},\ldots,x_{k})$ be the
constraint application of $B$ with constants that results if for all
$1 \leq i \leq k$, $x_{i} = $
\begin{itemize}
\item $1$ if $s_i = t_i = u_i = 1$,
\item $0$ if $s_i = t_i = u_i = 0$,
\item $x$ if $s_i = t_i = 0$ and $u_i = 1$,
\item $x'$ if $s_i = t_i = 1$ and $u_i = 0$,
\item $y$ if $s_i = u_i = 0$ and $t_i = 1$,
\item $y'$ if $s_i = u_i = 1$ and $t_i = 0$,
\item $z$ if $s_i = 0$ and $t_i = u_i = 1$,
\item $z'$ if $s_i = 1$ and $t_i = u_i = 0$.
\end{itemize}
Note that 
\begin{itemize}
\item $\widehat{B}(0,0,0,1,1,1) = B(s) = 1$,
\item $\widehat{B}(0,1,1,1,0,0) = B(t) = 1$, 
\item $\widehat{B}(1,0,1,0,1,0) = B(u) = 1$, 
\item $\widehat{B}(1,1,0,0,0,1) = B(s \oplus t \oplus u) = 1$,
\item and $\widehat{B}(0,0,1,1,1,0) = B(\majority(s,t,u)) = 0$.
\end{itemize}

Let $S = \{\widehat{B}(x,y,z,x',y',z'), A(x,x'),A(y,y'),A(z,z')\}$.
Then $S$ is a set of constraint applications of ${\cal C}$ with
constants such that there exists a ternary function $h$ such that
$S(x,y,z,x',y',z') = h(x,y,z) \wedge (x \oplus x') \wedge (y \oplus
y') \wedge (z \oplus z')$ and $h(000) = h(011) = h(101) = h(110) = 1$
and $h(001) = 0$.

The following table summarizes the possibilities we have.
\begin{center}
\begin{tabular}{|c|c|c|c|c|c|c|c|c|c|c|}
$xyz$&000&001&010&011&100&101&110&111\\
\hline
$h(x,y,z)$&1&0&$a$&1&$b$&1&1&$c$
\end{tabular}
\end{center}
We will analyze all cases.

\begin{itemize}
\item $a = 1$.  In this case, $S(0,y,z,1,y',z') =
(y \vee z') \wedge (y \oplus y') \wedge (z \oplus z')$,
and the result follows from Lemma~\ref{l:gitoorxor}.\ref{l:2}.
\item $b = 1$. In this case, $S(x,0,z,x',1,z') =
(x \vee z') \wedge (x \oplus x') \wedge (z \oplus z')$,
and the result follows from Lemma~\ref{l:gitoorxor}.\ref{l:2}.
\item $b = 0$ and $c = 1$. In this case, $S(1,y,z,0,y',z') =
(y \vee z) \wedge (y \oplus y') \wedge (z \oplus z')$,
and the result follows from Lemma~\ref{l:gitoorxor}.\ref{l:2}.
\item $a = b = c = 0$.  In this case,  
$S(x',y,z,x,y',z') = (x \oplus y \oplus z) \wedge (x \oplus x') \wedge
(y \oplus y') \wedge (z \oplus z')$, and the result follows
from Lemma~\ref{l:gitothreeaffine}.
\end{itemize}
\end{proof}

\section{Proof of Claim \ref{cl:rottennf}}
\label{proof:cl:rottennf}

Suppose for a contradiction that
$a$, $b$, and $c$ are three distinct variables in $X$
such that $U(\widehat{G}) \rightarrow \oneinthree(a,b,c)$ and
$\oneinthree(a,b,c) \not \in U(\widehat{G})$.

First note that that it cannot be the case that 
$\{a,b,c\}$ contains $\{x_i,x'_i\}$ or $\{y_i,y'_i\}$ for some $i$, since that
would imply that $U(\widehat{G}) \rightarrow \neg d$ for some 
variable $d \in X$. But clearly, there exists a satisfying assignment
for $U(\widehat{G})$ such that the value of $d$ is 1.

Secondly, note that if we set all $x'$-variables and all $y$-variables to $1$,
and all other variables in $X$ to 0, we obtain a 
satisfying assignment for $U(\widehat{G})$.
It follows that exactly one variable in
$\{a, b, c\}$ is an $x'$-variable or a $y$-variable.
The proof consists of a careful analysis of the different
cases.  We will show that in each case, there exists an assignment
that satisfies $U(\widehat{G})$  but that does
not satisfy $\oneinthree(a,b,c)$,
which contradicts the assumption that
$U(\widehat{G}) \rightarrow \oneinthree(a,b,c)$.

\begin{enumerate}
\item
If $\{a,b,c\} = \{x_i, x_j, y_k\}$, then,
since $\oneinthree(a,b,c) \not \in U(\widehat{G})$,
$e_k \neq \{i,j\}$.  Without loss of generality, let $j \not \in e_k$.
It is easy to see that there is a satisfying assignment for
$U(\widehat{G})$ such that $y_k$ and $x_j$ are set to 1.
(Set all other $x$-variables to 0 and set
$y_{\ell}$ to 1 iff $j \not \in e_{\ell}$.)
Thus, we have an assignment that satisfies $U(\widehat{G})$ but not
$\oneinthree(a,b,c)$. 
\item
If $\{y'_{\ell}, y_k\} \subseteq \{a,b,c\}$,
note that $k \neq \ell$, by the observation made
above.  Let $i$ be such that $i \in e_{\ell}$ and $i \not \in e_k$.
Set $x_i$ to 1, and set all other $x$-variables to $0$.
This can be extended to a satisfying assignment for $S(\widehat{U})$,
and in this assignment, $y_{\ell}$ is 0 (and thus $y'_{\ell}$ is 1), and
$y_k = 1$.
\item
If $\{x'_i, x_j\} \subseteq \{a,b,c\}$.
Then $i \neq j$.  Set $x_j$ to 1 and all other
$x$-variables to 0.  It is easy to see that this can be extended
to a satisfying assignment for $U(\widehat{G})$.
\item
If $\{a,b,c\} = \{x'_i, y'_k, y'_{\ell}\}$, then, 
if $i \not \in e_{\ell}$ and $i \not \in e_k$,
then set $x_i$ to $1$, set all other $x$-variables to $0$, set 
$y_r$ to 1 iff $i \not \in e_r$, and extend this to a satisfying
assignment for $U(\widehat{G})$.  If $i \in e_{\ell}$ or $i \in e_k$,
assume without loss of generality that $i \in e_k$,
set $x_i$ to 0, set $y_k$ to 0, and extend this to a satisfying
assignment for $U(\widehat{G})$.
\end{enumerate}

\section{Proof of Claim \ref{cl:maxxor}}
\label{proof:cl:maxxor}

Suppose that $a$ and $b$ are two distinct variables in $X$ such that
$U(\widehat{G}) \rightarrow (a \oplus b)$ and $a \oplus b \not \in
U(\widehat{G})$.  It is easy to see that we can set $a$ and $b$ to $0$,
and extend this to a satisfying assignment of $U(\widehat{G})$, which
is a contradiction.

Next, let $a, b, c$ be three distinct variables in $X$ 
such that $U(\widehat{G}) \rightarrow (a \oplus b \oplus c)$
and $a \oplus b \oplus c \not \in U(\widehat{G})$. 
Let $\widehat{X} = \{x_1, \ldots, x_n, y_1, \ldots, y_m\}$.
Any assignment that sets all variables in $\widehat{X}$ to 1, and
all variables not in $\widehat{X}$ to 0, satisfies $U(\widehat{G})$.
It follows that either exactly one or exactly three elements of $\{a,b,c\}$
are in $\widehat{X}$.

If exactly one of $\{a,b,c\}$ is in $\widehat{X}$, assume
without loss of generality that $a \in \widehat{X}$.
If $a' \in \{b,c\}$, then, without loss of generality, let
$a'=b$. In this case,  $U(\widehat{G}) \rightarrow \overline{b}$.
But this is a contradiction, since it is immediate that we
can set $c$ to 1 and extend this to a satisfying assignment of
$U(\widehat{G})$. Next, let $d, e \in \widehat{X}$ be such that
$d' = b$ and $e' = c$.  In that case, $a$, $d$, and $e$ are
distinct variables in $\widehat{X}$ such that
$U(\widehat{G}) \rightarrow a \oplus d \oplus e$
and $a \oplus d \oplus e \not \in U(\widehat{G})$.
This falls under the next case.

Finally, suppose that $a, b, c$ are three distinct variables in $\widehat{X}$
such that $U(\widehat{G}) \rightarrow a \xor b \xor c$ and
$a \xor b \xor c \not \in U(\widehat{G})$.
Let $\widehat{U}(\widehat{G}) =
\{(x_i \oplus x_j \oplus y_k) \condition e_k = \{i,j\}\}$, i.e.,
$\widehat{U}(\widehat{G})$ consists of all constraint applications
in $U(\widehat{G})$ whose variables are in $\widehat{X}$.
Since any assignment on $\widehat{X}$ that satisfies $\widehat{U}(\widehat{G})$
can be extended to a satisfying assignment of $U(\widehat{G})$
(by letting $a' = \overline{a}$ for all $a \in \widehat{X}$),
the desired result follows immediately from the following claim.

\begin{ourclaim}
\label{cl:maxxortwo}
$\widehat{U}(\widehat{G})$ is a maximal set
of constraint applications of $\XOR_3$ without duplicates
over variables $\widehat{X}$.
\end{ourclaim}

\begin{proof}
Suppose that $a, b, c$ are three distinct variables in $\widehat{X}$
such that $\widehat{U}(\widehat{G}) \rightarrow a \xor b \xor c$ and
$a \xor b \xor c \not \in \widehat{U}(\widehat{G})$.
The proof consists of a careful analysis of the different cases.
We will show that in each case, there exists an assignment 
on $\widehat{X}$ 
that satisfies $\widehat{U}(\widehat{G})$  but not $(a \xor b \xor c)$,
which contradicts the assumption that
$\widehat{U}(\widehat{G}) \rightarrow a \xor b \xor c$.

It is important to note that any assignment to $\{x_1, \ldots, x_n\}$
can be extended
to a satisfying assignment of $\widehat{U}(\widehat{G})$.

\begin{enumerate}
\item If $a$, $b$, and $c$ are in $\{x_1, \ldots, x_n\}$, then
set $a$, $b$, and $c$ to $0$. 
This assignment can be extended to an assignment
on $\widehat{X}$ that satisfies 
$x_i \xor x_j \xor y_k$ for $e_k = \{i,j\}$.
So, we now have an assignment that satisfies
$\widehat{U}(\widehat{G})$  but does not satisfy $(a \xor b \xor c)$.

\item If exactly two of $\{a,b,c\}$ are in $\{x_1, \ldots, x_n\}$, then
without loss of generality, let
$c = y_k$ for $e_k = \{i,j\}$.
By the assumption that $a \xor b \xor c$
is not in $\widehat{U}(\widehat{G})$, at least one of $a$ and $b$ is not
in $\{x_i,x_j\}$.

Without loss of generality, let $a \not \in
\{x_i,x_j\}$. Set $a$ to 0 and set $\{x_1, \ldots, x_m\} \setminus\{a\}$ to $1$.
This assignment can be extended to a satisfying assignment for
$\widehat{U}(\widehat{G})$. Note that such an assignment will set $y_k$ to $1$.
It follows that this assignment does not satisfy $a \xor b \xor c$. 

\item If exactly one of $\{a,b,c\}$ are in $\{x_1, \ldots, x_n\}$,
without loss of generality, let $a \in \{x_1, \ldots, x_n\}$.
Set $a$ to $0$ and $b$ and $c$ to $1$. It is easy to see that
this can be extended to a satisfying assignment for $\widehat{U}(\widehat{G})$.

\item If  $a$, $b$, and $c$ are in $\{y_1, ..., y_m\}$, let $a = y_{k_1}, b = y_{k_2},
c = y_{k_3}$ such that $e_{k_\ell} = \{i_{\ell},j_{\ell}\}$
for $\ell \in \{1,2,3\}$.
First suppose that for every $\ell \in \{1, 2, 3\}$, for every
$x \in \{x_{i_{\ell}},x_{j_{\ell}}\}$, there exists an
$\ell' \in \{1,2,3\}$ with  $\ell' \neq \ell$ and a constraint
application $A$ in $\widehat{U}(\widehat{G})$ such that $x$ and $y_{k_{\ell'}}$ occur in $A$. 
This implies that every vertex in 
$\{i_1, j_1, i_2, j_2, i_3, j_3\}$ is incident with at least
2 of the edges in $e_{k_1}, e_{k_2}, e_{k_3}$.
Since these three edges are distinct, it
follows that the edges $e_{k_1}, e_{k_2}, e_{k_3}$ form a triangle
in $\widehat{G}$, which contradicts the assumption that $\widehat{G}$
is triangle-free.

So, let $\ell \in \{1,2,3\}$, $x \in \{x_{i_{\ell}},x_{j_{\ell}}\}$
be such that for all $\ell' \in \{1,2,3\}$ with $\ell \neq \ell'$,
$x$ and $y_{k_{\ell'}}$ do not occur in the same constraint application 
in $\widehat{U}(\widehat{G})$.  Set $x$ to $0$ and set $\{x_1, \ldots, x_n\}
\setminus \{x\}$ to $1$. This can be
extended to a satisfying assignment of $\widehat{U}(\widehat{G})$ and such a 
satisfying assignment must have the property that $y_{k_{\ell}}$ is $0$ and
$y_{k_{\ell'}}$ is $1$ for all $\ell' \in \{1,2,3\}$ such that
$\ell' \neq \ell$.
\end{enumerate}
\end{proof}

\section{Removing Constants}\label{s:rem-const}

In Section~\ref{s:gihard}, we showed that for all ${\cal C}$ that
are not 2-affine, $\GI \manyone \ISO_c({\cal C})$.  In this section,
we will show that we can get the same result without using
constants.

\begin{theorem}
\label{t:noconst}
If ${\cal C}$ is not 2-affine, then 
$\GI \leq_m^p \ISO({\cal C})$.
\end{theorem}

This completes the proof of Theorem~\ref{t:isotrich}.

Note that constants are used a lot in the proofs in the
previous section.  It is known that removing constants
can be a lot of work, see, for example Dalmau's work to remove
constants from quantified constraint applications~\cite{dalmau-tr}.

Part of the problem is that there are far more cases to consider
than in the case with constants.  Recall that in the case with constants,
if sufficed to prove GI-hardness for 6 constraints
(namely, the constraints from Lemma~\ref{l:gitoor} and
Lemma~\ref{l:gitothreeaffine}), since
it follows from the proofs of Theorems~\ref{th:gitonotaffine}
and~\ref{th:gitonotbijunctive} that
for all ${\cal C}$ that are not 2-affine, there exists
a set of constraint applications with constants that is 
equivalent to one of these 6 constraints.  Now look at one
of the simplest of these 6 constraints, namely, $\OR_0$.
If there exists a set of constraint applications $S(0,1,x,y)$
of ${\cal C}$ such that  $S(0,1,x,y) \equiv \OR_0(x,y)$,
and we can use constants, it suffices to show GI-hardness
for $\OR_0$.  But if we cannot use constants, we need to show
GI-hardness for all $2^{12}$ constraints $A$ 
of arity 4 such that $A(0,1,x,y) \equiv \OR_0$. 
For the most complicated of the 6 constraints from the previous section,
which has arity 6, we will now have $2^{192}$
cases to consider.  Clearly, handling
each of these constraints separately is not an option.

We need a way to uniformly transform the GI-hardness reductions
for the case with constants from the previous section into
GI-hardness reductions to the corresponding constant-free
isomorphism problems.

A crucial tool in this transformation will be the following lemma,
Lemma~\ref{l:noconst}.  This lemma does not immediately
imply the result, but it will enable us to transform the GI-hardness
reductions from the previous section into GI-hardness reductions 
to the corresponding case without constants.

Say that a constraint $C$ is {\em complementative} (or {\em
C-closed}) if for every $s \in \{0,1\}^k$, $C(s) = C(\overline{s})$,
where $k$ is the arity of $C$ and $\overline{s} \in \{0,1\}^k \eqd
\vec{1} - s$, i.e., $\overline{s}$ is obtained by flipping every bit
of $s$. A set of constraints is complementative if each of its elements is.

Often, we will want to be explicit about the variables and/or
constants that occur in a set of constraint applications (with constants).
In such cases, we will write (sets of) constraint applications as 
$S(x_1,\ldots,x_k)$ or $S(0,1,x_1,\ldots,x_k)$.
It is important to recall that in our terminology,
a set of constraint applications of ${\cal C}$ does
not contain constants.

\begin{lemma}
\label{l:noconst}
If ${\cal C}$ is not 2-affine, then
\begin{itemize}
\item  there exists
a set $S(x,y)$ of constraint applications of ${\cal C}$
such that $S(x,y)$ is equivalent  to
$\overline{x} \wedge y$, 
$\overline{x} \vee y$, 
$x \oplus y$, or
$x \leftrightarrow y$; or
\item
there exists a set $S(t,x,y)$ of constraint applications
of ${\cal C}$ such that $S(t,x,y)$ is equivalent to
$t \wedge (\overline{x} \vee y)$, 
$t \wedge (x \leftrightarrow y)$, or
$t \wedge (x \vee y)$; or
\item
there exists a set $S(f,x,y)$ of constraint applications
of ${\cal C}$ such that $S(f,x,y)$ is equivalent to
$\overline{f} \wedge (\overline{x} \vee y)$, 
$\overline{f} \wedge (x \leftrightarrow y)$, or
$\overline{f} \wedge (\overline{x} \vee \overline{y})$.
\end{itemize}
\end{lemma}

\begin{proof}
Let $A(0,1,x,y)$ be a constraint application of
${\cal C}$ with constants such that $A(0,1,x,y)$ is equivalent  to
$\OR_0(x,y)$, $\OR_1(x,y)$, $\OR_2(x,y)$, or $\XOR_2(x,y)$.
The existence of $A(0,1,x,y)$ follows immediately from
the proof of Theorem~\ref{th:gitonotaffine} and from
the first paragraph of the proof of Theorem~\ref{th:gitonotbijunctive}.

As is usual in proofs of dichotomy theorems for Boolean constraints,
our proof uses case distinctions. 
The main challenge of the proof is to
keep the number of cases in check.

\begin{description}
\item[$A$ is not 0-valid, not 1-valid, and not complementative]

In this case, there exists a constraint application
of $A$ that is equivalent to $\overline{x} \wedge y$~\cite{CrHe96},
see also \cite[Lemma~5.25]{CrKhSu00}.

\item[$A$ is 0-valid, 1-valid, and not complementative]

In this case, there exists a constraint application
of $A$ that is equivalent to $\overline{x} \vee y$~\cite[Lemma~4.13]{CrHe96},
see also \cite[Lemma~5.25]{CrKhSu00}.

\item[$A$ is not 0-valid, not 1-valid, and complementative]

In this case, there exists a constraint application 
of $A$ that is equivalent to $x \xor y$~\cite{CrKhSu00}, proof of Lemma~5.24.

\item[$A$ is 0-valid, 1-valid, and complementative]

Let $\alpha$ be an assignment that does not satisfy $A(x_1, x_2, x_3, x_4)$.
Such an assignment exists, since $A(x_1, x_2,\allowbreak x_3, x_4)$
is not a tautology.
Let $B(x,y)$ be the constraint application of $A$ that results
when replacing all variables in $A(x_1, x_2, x_3, x_4)$ whose
value is true in $\alpha$ by $x$, and 
replacing all variables in $A(x_1, x_2, x_3, x_4)$
whose value is false in $\alpha$
by $y$.  Note that $B(0,0) = B(1, 1)  = 1 $, since 
$B$ is 0-valid and 1-valid;  $B(1,0) = 0$ by construction;
and $B(0,1) = 0$, since $B$ is complementative.
It follows that $B$ is equivalent to $x \leftrightarrow y$.

\item[$A$ is 1-valid and not 0-valid]

In this case we can force a variable $z$ to be true
by adding the constraint application $A(z,z,z,z)$.

Let $B$ be the 3-ary constraint such that $B(f,x,y) \equiv A(f,1,x,y)$.
Then $B$ is 1-valid, since $A$ is 1-valid.  If $B$ is 0-valid,
then, by the cases handled above, there exists a constraint
application of $B$ that is equivalent to $\overline{x} \vee y$ or
to $x \leftrightarrow y$.
Since $\{A(f,t,x,y), A(t,t,t,t)\}$ is equivalent to 
$t \wedge B(f,x,y)$, 
it follows that there exists a set of constraint applications
of $A$ that is equivalent to $t \wedge (\overline{x} \vee y)$ or to
$t \wedge (x \leftrightarrow y)$.

It remains to handle the case that $B$ is not 0-valid, i.e.,
the case that $A(0,1,0,0) = 0$.  Since $A(0,1,0,0) = 0$,
we have that
$A(0,1,x,y) \equiv \OR_0(x,y)$ or $A(0,1,x,y) \equiv \XOR_2(x,y)$.

First consider the case that $A(0,1,x,y) \equiv \OR_0(x,y)$.

If $A(1,1,0,1) = 0$, then consider the set
\[S(t,x,y) = \{A(x,t,y,t), A(t,t,t,t)\}.\]
This set is equivalent to $t \wedge (\overline{x} \vee y)$,
since $A(0,1,0,1) = 1$;
$A(0,1,1,1) = 1$;
$A(1,1,0,1) = 0$; and
$A(1,1,1,1) = 1$.

If $A(1,1,0,1) = 1$, then consider the set
\[S(t,x,y) = \{A(x,t,y,x), A(t,t,t,t)\}.\]
This set is equivalent to $t \wedge (x \vee y)$,
since $A(0,1,0,0) = 0$;
$A(0,1,1,0) = 1$;
$A(1,1,0,1) = 1$; and
$A(1,1,1,1) = 1$.

Finally, consider the case that $A(0,1,x,y) \equiv \XOR_2(x,y)$.
We are in the
following situation:
$A(0,1,0,0) = 0, A(0,1,0,1) = 1, A(0,1,1,0) = 1, A(0,1,\allowbreak 1,1) = 0$, and
$A(1,1,1,1) = 1$.  Consider the following set
\[S(t,x,y) = \{A(t,t,t,t), A(x,t,y,t)\}.\]
If $A(1,1,0,1) = 1$, then $S(t,x,y) \equiv t \wedge (x \vee \overline{y})$.
If $A(1,1,0,1) = 0$, then $S(t,x,y) \equiv t \wedge (x \leftrightarrow y)$.

\item[$A$ is 0-valid and not 1-valid]

In this case we can force a variable $z$ to be false
by adding the constraint application $A(z,z,z,z)$.

Let $B$ be the 3-ary constraint such that $B(t,x,y) \equiv A(0,t,x,y)$.
Then $B$ is 0-valid, since $A$ is 0-valid.  If $B$ is 1-valid,
then, by the cases handled above, there exists a constraint
application of $B$ that is equivalent to $\overline{x} \vee y$ or
to $x \leftrightarrow y$.
Since $\{A(f,t,x,y), A(f,f,f,f)\}$ is equivalent to 
$\overline{f} \wedge B(f,x,y)$, 
it follows that there exists a set of constraint applications
of $A$ that is equivalent to $\overline{f} \wedge (\overline{x} \vee y)$ or to
$\overline{f} \wedge (x \leftrightarrow y)$.

It remains to handle the case that $B$ is not 1-valid, i.e.,
the case that $A(0,1,1,1) = 0$.  Since $A(0,1,1,1) = 0$,
$A(0,1,x,y) \equiv \OR_2(x,y)$.

If $A(0,0,1,0) = 0$, then consider the set
\[S(f,x,y) = \{A(f,x,y,f), A(f,f,f,f)\}.\]
This set is equivalent to $\overline{f} \wedge (x \vee \overline{y})$,
since $A(0,0,0,0) = 1$;
$A(0,1,0,0) = 1$;
$A(0,0,1,0) = 0$; and
$A(0,1,1,0) = 1$.

If $A(0,0,1,0) = 1$, then consider the set
\[S(f,x,y) = \{A(f,x,y,x), A(f,f,f,f)\}.\]
This set is equivalent to $\overline{f} \wedge
(\overline{x} \vee \overline{y})$,
since $A(0,0,0,0) = 1$;
$A(0,0,1,0) = 1$;
$A(0,1,0,1) = 1$; and
$A(0,1,1,1) = 0$.
\end{description}
\end{proof}

When we look at Lemma~\ref{l:noconst}, some of the cases
will give us the required reduction immediately or almost immediately.

\begin{lemma}
\label{l:gitonoconst}
$\GI$ is polynomial-time many-one reducible to
\begin{enumerate}
\item
$\ISO(\{\lambda xy.(\overline{x} \vee y)\})$,
\item
$\ISO(\{\lambda txy. (t \wedge (\overline{x} \vee y))\})$,
\item
$\ISO(\{\lambda txy. (t \wedge ({x} \vee y))\})$,
\item
$\ISO(\{\lambda fxy. (\overline{f} \wedge (\overline{x} \vee y))\})$,  and
\item
$\ISO(\{\lambda fxy. (\overline{f} \wedge (\overline{x} \vee \overline{y}))\})$.
\end{enumerate}
\end{lemma}

\begin{proof}
The first case follows from Lemma~\ref{l:gitoor}. For
the remaining cases, we will adapt the reductions of
of the proof of Lemma~\ref{l:gitoor}.

Let $\widehat{G}$ be a graph and let
$V(\widehat{G}) = \{1, 2, \ldots , n\}$.
Let $E(\widehat{G}) = \{e_1, \ldots, e_m\}$. 

Define
\begin{eqnarray*}
S_2(\widehat{G}) & = &
\{t \wedge (x_i \vee \overline{y_k}), t \wedge (x_j \vee \overline{y_k})
\condition e_k = \{i,j\}\}\\
S_3(\widehat{G})  & = & \{t \wedge (x_i \vee x_j) 
\condition \{i,j\} \in E(\widehat{G})\}\\
S_4(\widehat{G}) & = &
\{\overline{f} \wedge (x_i \vee \overline{y_k}),
\overline{f} \wedge (x_j \vee \overline{y_k})
\condition e_k = \{i,j\}\}\\
S_5(\widehat{G}) & = &
\{\overline{f} \wedge (\overline{x_i} \vee \overline{x_j})
\condition \{i,j\} \in E(\widehat{G})\}
\end{eqnarray*}

Let $G$ and $H$ be two graphs without
isolated vertices and with vertex set $\{1, 2, \ldots, n\}$.
We will show that for $2 \leq i \leq 5$,
$G$ is isomorphic to $H$ if and only if
$S_i(G)$ is  isomorphic to $S_i(H)$.

For $i = 2,3$, it follows from the
proof of Lemma~\ref{l:gitoor}
that $G$ is isomorphic to $H$ if and only if
$S_i(G)[t := 1]$ is isomorphic to $S_i(H)[t := 1]$.

Note that $S_i(G) \equiv (S_i(G)[t := 1] \wedge t)$
and  $S_i(H) \equiv (S_i(H)[t := 1] \wedge t)$.
Thus, if $S_i(G)[t := 1]$ is isomorphic to $S_i(H)[t := 1]$,
then $S_i(G)$ is isomorphic to $S_i(H)$.

For the converse, note that
if $S_i(G)$ is isomorphic to $S_i(H)$, then the isomorphism
must map $t$ to $t$, since $t$ is the unique variable $z$ such
that $S_i(G) \rightarrow z$ and the unique variable
$z$ such that $S_i(H) \rightarrow z$.
It follows immediately 
that $S_i(G)[t := 1]$ is isomorphic to $S_i(H)[t := 1]$,
by the same isomorphism.

For $i = 4,5$, it follows from the
proof of Lemma~\ref{l:gitoor}
that $G$ is isomorphic to $H$ if and only if
$S_i(G)[f := 0]$ is isomorphic to $S_i(H)[f := 0]$.

Note that 
$S_i(G) \equiv (S_i(G)[f := 0] \wedge \overline{f})$
and  $S_i(H) \equiv (S_i(H)[f := 0] \wedge \overline{f})$.
Thus,
if $S_i(G)[f := 0]$ is isomorphic to $S_i(H)[f := 0]$,
then $S_i(G)$ is isomorphic to $S_i(H)$.

For the converse, note that
if $S_i(G)$ is isomorphic to $S_i(H)$, then the isomorphism
must map $f$ to $f$, since $f$ is the unique variable $z$ such
that $S_i(G) \rightarrow \overline{z}$ and the unique variable
$z$ such that $S_i(H) \rightarrow \overline{z}$.
It follows immediately 
that $S_i(G)[f := 0]$ is isomorphic to $S_i(H)[f := 0]$,
by the same isomorphism.
\end{proof}

To complete the proof of Theorem~\ref{t:noconst},
it remains to show the following claim.
\begin{ourclaim}
\label{cl:transform}
If ${\cal C}$ is not 2-affine, and
\begin{itemize}
\item there exists
a set $U(x,y)$ of constraint applications of ${\cal C}$
such that $U(x,y)$ is equivalent  to
$\overline{x} \wedge y$, 
$x \oplus y$, or
$x \leftrightarrow y$; or
\item
there exists a set $U(t,x,y)$ of constraint applications
of ${\cal C}$ such that $U(t,x,y)$ is equivalent to
$t \wedge (x \leftrightarrow y)$, or
\item 
 there exists a set $U(f,x,y)$ of constraint applications
of ${\cal C}$ such that $U(f,x,y)$ is equivalent to
$\overline{f} \wedge (x \leftrightarrow y)$,
\end{itemize}
then $\GI \manyone \ISO({\cal C})$.
\end{ourclaim}

These cases will be handled by transforming the GI-hardness reductions
from the previous section into GI-hardness reductions to the corresponding
constant-free isomorphism problems using the set of constraint
applications $U$ from Claim~\ref{cl:transform}.

We first restate some of the definitions and results
from the previous section in a way that is explicit
about the occurrences of constants.
 
\begin{definition}
\begin{enumerate}
\item
${\cal D}_1$ is the set of 4-ary constraints $D$ such that
$D(0,1,x,y) \equiv \OR_0(x,y)$.
\item
${\cal D}_2$ is the set of 4-ary constraints $D$ such that
$D(0,1,x,y) \equiv \OR_1(x,y)$.
\item
${\cal D}_3$ is the set of 4-ary constraints $D$ such that
$D(0,1,x,y) \equiv \OR_2(x,y)$.
\item
${\cal D}_4$ is the set of 6-ary constraints $D$ such that
$D(0,1,x,y,x',y') \equiv (x \vee y) \wedge (x
\oplus x') \wedge (y \oplus y')$.
\item 
${\cal D}_5$ is the set of 8-ary constraints $D$ such that
$D(0,1,x,y,z,\allowbreak x',y',z') \equiv \oneinthree(x,y,z)
\wedge (x \oplus x') \wedge (y \oplus y') \wedge (z \oplus z')$. 
\item
${\cal D}_6$ is the set of 8-ary constraints $D$ such that
$D$ is affine and
$D(0,1,x,y,z,\allowbreak x',y',z') \equiv (x
\oplus y \oplus z) \wedge (x \oplus x') \wedge (y \oplus y') \wedge (z
\oplus z')$.
\end{enumerate}
\end{definition}

{From} the previous section, we know the following fact.

\begin{fact}
\begin{itemize}
\item If ${\cal C}$ is not 2-affine, then there exists
a set of constraint applications $S(x_1, \ldots, x_k)$ of
${\cal C}$ such that  $S(x_1, \ldots, x_k) \equiv
D(x_1, \ldots, x_k)$ for some $D \in \bigcup_{1 \leq i \leq 6}
{\cal D}_i$.
\item For $1 \leq i \leq 6$, and for all $D \in {\cal D}_i$,
$\GI \manyone \ISO_c(\{D\})$.
\end{itemize}
\end{fact}

\begin{proof}
The second part follows immediately from Lemmas~\ref{l:gitoor}
and~\ref{l:gitothreeaffine}.  For the first part, it follows 
from the proofs of Theorems~\ref{th:gitonotaffine}
and~\ref{th:gitonotbijunctive}, that there exist
an $i$ such that $1 \leq i \leq 6$,
a constraint $D' \in {\cal D}_i$, and a set
$S(0,1,x_3, \ldots, x_k)$ of constraint applications of
${\cal C}$ with constants such that
$S(0,1,x_3, \ldots, x_k)$ is equivalent to $D'(0,1,x_3, \ldots, x_k)$.
By definition of ${\cal D}_i$, 
there exists a constraint $D \in {\cal D}_i$ such that
$S(x_1,x_2,x_3, \ldots, x_k)$ is equivalent to
$D(x_1,x_2,x_3, \ldots, x_k)$.
\end{proof}

Our goal is to remove the constants and
show that for $1 \leq i \leq 6$, and for all $D \in {\cal D}_i$,
$\GI \manyone \ISO(\{D\})$.

In order to do this, we will transform the reduction for the
case with constants into a reduction for the case without
constants.  We will need some properties from the case 
with constants.

\begin{definition}
Let $\widehat{G}$ be a graph.
\begin{enumerate}
\item
For $D \in {\cal D}_1$,
$S_{1,D}(\widehat{G}) =
\{D(f,t,x_i,x_j) \condition \{i,j\} \in E(\widehat{G})\}$.
\item
For $D \in {\cal D}_2$,
$S_{2,D}(\widehat{G}) = \{D(f,t,y_k,x_i), D(f,t,y_k,x_j)
\condition e_k = \{i,j\}\}$.
\item
For $D \in {\cal D}_3$,
$S_{3,D}(\widehat{G}) = \{D(f,t,x_i,x_j) \condition
\{i,j\} \in E(\widehat{G})\}$.
\item
For $D \in {\cal D}_4$,
$S_{4,D}(\widehat{G}) = \{D(f,t,x_i,x_j,x_i',x_j')
\condition \{i,j\} \in E(\widehat{G})\}$.
\item 
For $D \in {\cal D}_5$,
$S_{5,D}(\widehat{G}) =
\{D(f,t,x_i,x_j,y_k,x_i',x_j',y_k') \condition e_k = \{i,j\}\}$.
\item
For $D \in {\cal D}_6$,
$S_{6,D}(\widehat{G}) =
\{D(f,t,x_i,x_j,y_k,x_i',x_j',y_k') \condition e_k = \{i,j\}\}$.
\end{enumerate}
\end{definition}

{From} the proofs of Lemmas~\ref{l:gitoor}
and~\ref{l:gitothreeaffine}, 
we have the following fact, which
witnesses the GI-hardness for the case with constants.

\begin{fact}
\label{f:gitoconst}
Let $G$ and $H$ be graphs such that
$V(G) = V(H) = \{1, \ldots, n\}$,
$E(G) = \{e_1, \ldots, e_m\}$,
$E(H) = \{e'_1, \ldots, e'_m\}$,
all vertices in $G$ and $H$ have degree at least two,
and $G$ and $H$ do not contain triangles.

For all $1 \leq i \leq 6$ and for all $D \in {\cal D}_i$,
$G$ is isomorphic to $H$  if and only
if $S_{i,D}(G)[f := 0, t:= 1]$ is isomorphic to
$S_{i,D}(H)[f := 0, t:= 1]$.
\end{fact}

The following simple observation is also useful.  
\begin{observation}
\label{o:}
Let $G$ and $H$ be graphs such that
$V(G) = V(H) = \{1, \ldots, n\}$,
$E(G) = \{e_1, \ldots, e_m\}$,
$E(H) = \{e'_1, \ldots, e'_m\}$,
all vertices in $G$ and $H$ have degree at least two,
and $G$ and $H$ do not contain triangles.

For all $1 \leq i \leq 6$ and for all $D \in {\cal D}_i$,
if $G$ is isomorphic to $H$, then 
$S_{i,D}(G)$ is isomorphic to
$S_{i,D}(H)$ by an isomorphism that maps $f$ to $f$ and
$t$ to $t$.
\end{observation}

\begin{proof}
Let $\pi$ be an isomorphism from $G$ to $H$.
Then $S_{i,D}(G)$ is isomorphic to $S_{i,D}(H)$, by an isomorphism
that maps $x_i$ to $x_{\pi(i)}$, 
$x'_i$ to $x'_{\pi(i)}$, 
$y_k$ to $y_{\ell}$ and
$y'_k$ to $y'_{\ell}$ for $e_k = \{i,j\}$, $e'_{\ell} = \{\pi(i),\pi(j)\}$,
$f$ to $f$,  and $t$ to $t$.  Note that this isomorphism even
makes the sets of constraints equal, rather than merely equivalent.
\end{proof}

In order to remove the constants, we need to use the set
$U$ from Claim~\ref{cl:transform} as well as certain
properties of $S_{i,D}$.

\begin{lemma}
\label{l:copies}
Let $\widehat{G}$ be a graph such that
$V(\widehat{G}) = \{1, \ldots, n\}$,
$E(\widehat{G}) = \{e_1, \ldots, e_m\}$,
all vertices in $\widehat{G}$ have degree at least two,
and $\widehat{G}$ does not contain triangles.
Let $1 \leq i \leq 6$, let $D \in {\cal D}_i$,
and let $z$ and $z'$ be any two distinct variables. If
\[S_{i,D}(\widehat{G}) \cup \{\overline{f},
\overline{f_1}, \overline{f_2}, t, t_1\} \rightarrow
(z \leftrightarrow z'),\]
then $\{z,z'\} = \{f, f_1\}$, $\{f, f_2\}$, $\{f_1, f_2\}$, or $\{t, t_1\}$.
\end{lemma}

\begin{proof}
If $S_{i,D}(\widehat{G}) \cup \{\overline{f},
\overline{f_1}, \overline{f_2}, t, t_1\} \rightarrow
(z \leftrightarrow z')$, then
$S_{i,D}(\widehat{G}) \cup \{\overline{f},
\overline{f_1}, \overline{f_2}, t, t_1\}\allowbreak[f, f_1, f_2 := 0, t, t_1 := 1] 
\rightarrow (z \leftrightarrow z')[f, f_1, f_2 := 0, t, t_1 := 1]$.
Since $f_1, f_2$, and $t_1$ do not occur in 
$S_{i,D}(\widehat{G})$,  this implies that
$S_{i,D}(\widehat{G})[f := 0, t := 1] 
\rightarrow (z \leftrightarrow z')[f, f_1, f_2 := 0, t, t_1 := 1]$.

Let $S(\widehat{G}) =  S_{i,D}(\widehat{G})[f := 0, t := 1].$
It suffices to show that
$S(\widehat{G}) \not \rightarrow
(z \leftrightarrow z')[f, f_1, f_2 := 0, t, t_1 := 1]$
for all pairs of distinct variables $z$ and $z'$ such that
$\{z,z'\} \not \in  \{\{f, f_1\}, \{f, f_2\}, \{f_1, f_2\}, \{t, t_1\}\}$.
Since $S(\widehat{G})$ is satisfiable, it is immediate that
$S(\widehat{G}) \not \rightarrow
(z \leftrightarrow z')[f, f_1, f_2 := 0, t, t_1 := 1]$
for $z \in \{f, f_1, f_2\}$ and $z' \in \{t, t_1\}$ or vice-versa.

It is easy to see that it remains to show that
for all variables $z$,
\begin{enumerate}
\item
$S(\widehat{G}) \not \rightarrow z$,  i.e., 
$S(\widehat{G})[z := 0]$ is satisfiable.
\item
$S(\widehat{G}) \not \rightarrow \overline{z}$, i.e.,
$S(\widehat{G})[z := 1]$ is satisfiable.
\item
For all variables $z' \neq z$,  
$S(\widehat{G}) \not \rightarrow (z \leftrightarrow z')$, i.e.,
$S(\widehat{G})[z := 1, z' := 0]$ is satisfiable or
$S(\widehat{G})[z := 0, z' := 1]$ is satisfiable.
\end{enumerate}

Since the definition of $S_{\hat{\imath},D}$ is different for each value of
$\hat{\imath}$, we will handle the different values of $\hat{\imath}$
separately.

\begin{enumerate}
\item  In this case, 
$S(\widehat{G})$ is equivalent to
$\{x_i \vee x_j \condition \{i,j\} \in E(\widehat{G})\}$.

For all $i$,  and all $j \neq i$,
$S(\widehat{G})[x_i := 1]$,
$S(\widehat{G})[x_i := 0]$, and
$S(\widehat{G})[x_i := 0, x_j := 1]$ are satisfiable
by setting all (remaining) variables to 1.

\item
In this case,
$S(\widehat{G})$ is equivalent to $\{x_i \vee \overline{y_k},
x_j \vee \overline{y_k} \condition e_k = \{i,j\}\}$.

For all $i$ and $k$,
$S(\widehat{G})[x_i := 1]$,
$S(\widehat{G})[y_k := 0]$, and
$S(\widehat{G})[x_i := 1, y_k := 0]$ are satisfiable
by setting all $x$-variables to 1 and all $y$-variables to 0.
For all $i$ and all $j \neq i$,
$S(\widehat{G})[x_i := 0]$, and
$S(\widehat{G})[x_i := 0, x_j := 1]$ are satisfiable
by setting all $y$-variables to 0.
For all $k$ and all $\ell \neq k$,
$S(\widehat{G})[y_k := 1]$, and
$S(\widehat{G})[y_k := 0, y_{\ell} := 1]$ are satisfiable
by setting all $x$-variables to 1.

\item
In this case,
$S(\widehat{G})$ is equivalent to
$\{\overline{x_i} \vee \overline{x_j} \condition
\{i,j\} \in E(\widehat{G})\}$.

For all $i$,  and all $j \neq i$,
$S(\widehat{G})[x_i := 1]$,
$S(\widehat{G})[x_i := 0]$, and
$S(\widehat{G})[x_i := 0, x_j := 1]$ are satisfiable
by setting all variables to 0.

\item
In this case, $S(\widehat{G})$ is equivalent to
$\{(x_i \vee x_j) \wedge (x_i \oplus  x_i') \wedge (x_j \oplus x_j')
\condition \{i,j\} \in E(\widehat{G})\}$.

For all $i,j$,
$S(\widehat{G})[x_i := 1]$,
$S(\widehat{G})[x'_j := 0]$, and
$S(\widehat{G})[x_i := 1, x'_j := 0]$ are satisfiable
by setting all $x$-variables  to 1 and all $x'$-variables to 0.

For all $i$ and all $j \neq i$, 
$S(\widehat{G})[x_i := 0]$ and
$S(\widehat{G})[x_i := 0, x_j := 1]$ are
satisfiable by setting all $x$-variables to 1,
$x'_i$ to 1, and all other $x'$-variables to 0.
$S(\widehat{G})[x'_i := 1]$ and
$S(\widehat{G})[x'_i := 1, x'_j := 0]$ are
satisfiable by setting all $x'$-variables to 0,
$x_i$ to 0 and all other $x$-variables to 1.

\item 
In this case, $S(\widehat{G})$ is equivalent to
$\{\oneinthree(x_i,x_j,y_k) \wedge (x_i \oplus x_i') \wedge (x_j \oplus x_j')
\wedge (y_k \oplus y_k') \condition e_k = \{i,j\}\}$.

For all $i,j, k, \ell$,
$S(\widehat{G})[x_i := 0]$,
$S(\widehat{G})[x'_j := 1]$,
$S(\widehat{G})[y_k := 1]$,
$S(\widehat{G})[y'_{\ell} := 0]$,
$S(\widehat{G})[x_i := 0, x'_j := 1]$,
$S(\widehat{G})[y_k := 1, y'_{\ell} := 0]$,
$S(\widehat{G})[x_i := 0, y_k := 1]$, and
$S(\widehat{G})[x'_j := 1, y'_{\ell} := 0]$, are all
satisfiable by by setting all $x$-variables and $y'$-variables to 0,
and all $y$-variables and $x'$-variables to 1.

For all $i$, and for all $j \neq i$,
$S(\widehat{G})[x_i := 1]$,
$S(\widehat{G})[x'_i := 0]$,
$S(\widehat{G})[x_i := 1, x_j := 0]$, and
$S(\widehat{G})[x'_i := 0, x'_j := 1]$ are satisfiable
by setting $x_i$ to 1 and all other $x$-variables to 0,
setting  $x'_i$ to 0 and all other $x'$-variables to 1,
setting $y_k$ to 0 and $y'_k$ to 1
for all $k$ such that $i \in e_k$, 
and setting all other $y$-variables to 1
and all other $y'$-variables to 0.

For all $k$, and for all $\ell \neq k$,
$S(\widehat{G})[y_k := 0]$,
$S(\widehat{G})[y'_k := 1]$,
$S(\widehat{G})[y_k := 0, y_\ell := 1]$, and
$S(\widehat{G})[y'_k := 1, y'_\ell := 0]$ are satisfiable:
Let $i$ be such that $i \in e_k$ and $i \not \in e_\ell$.
Set $x_i$ to 1, $x'_i$ to 0,
set all other $x$-variables to 0, and all other $x'$-variables to 1.
For all $k'$ such that $i \in e_{k'}$, set $y_{k'}$ to 0
and $y'_{k'}$ to 1, and set all other $y$-variables to 1
and all other $y'$-variables to 0.  Note that $y_{\ell}$ will be set
to 1, since $i \not \in e_{\ell}$.

For all $i$ and all $k$,
$S(\widehat{G})[x'_i := 1, y_k := 0]$ and
$S(\widehat{G})[x'_i := 0, y'_k := 1]$  are satisfiable:
Let $j$ be such that $e_k = \{i,j\}$.  Set $x_j$ to 1,
$x'_j$ to 0, set all other $x$-variables to 0,
all other $x'$-variables to 1, 
set $y_\ell$ to 0 and $y'_\ell$ to 1
for all $\ell$ such that $j \in e_\ell$, 
and set all other $y$-variables to 1
and all other $y'$-variables to 0.

\item
In this case, $S(\widehat{G})$ is equivalent to
$\{(x_i \oplus x_j \oplus y_k)
\wedge (x_i \oplus x_i') \wedge (x_j \oplus x_j')
\wedge (y_k \oplus y_k') \condition e_k = \{i,j\}\}$.

Note that
$\{\oneinthree(x_i,x_j,y_k) \wedge (x_i \oplus x_i') \wedge (x_j \oplus x_j')
\wedge (y_k \oplus y_k') \condition e_k = \{i,j\}\}
\rightarrow S(\widehat{G})$. Thus, the required properties
follow immediately from the previous case.
\end{enumerate}
\end{proof}

Finally, we are ready to prove Claim~\ref{cl:transform} which completes the
proof of Theorem~\ref{t:noconst}.

\begin{prooftext}{Claim~\ref{cl:transform}}
Let ${\cal C}$ be not 2-affine.
Let $U(X)$ be a set of constraint applications of ${\cal C}$ fulfilling
the statement of Claim~\ref{cl:transform}, i.e.,
\begin{itemize}
\item $U(x,y)$ is equivalent  to
$\overline{x} \wedge y$, 
$x \oplus y$, or
$x \leftrightarrow y$; or
\item
$U(t,x,y)$ is equivalent to $t \wedge (x \leftrightarrow y)$, or
\item $U(f,x,y)$ is equivalent to
$\overline{f} \wedge (x \leftrightarrow y)$.
\end{itemize}
Let $D$ be a constraint and $1 \leq i \leq 6$ be
such that $D \in {\cal D}_i$ and there exists a set
of constraint applications $S(x_1, \ldots, x_k)$ of
${\cal C}$ such that  $S(x_1, \ldots, x_k) \equiv
D(x_1, \ldots, x_k)$.  For $\widehat{G}$ a graph, we will
write $S(\widehat{G})$ for $S_{i,D}(\widehat{G})$.

Let $\widehat{G}$ be a graph such that
$V(\widehat{G}) = \{1, \ldots, n\}$ and
$E(\widehat{G}) = \{e_1, \ldots, e_m\}$.
For each of the cases for $U$, we will define
a polynomial-time computable set
$\widehat{S}(\widehat{G})$ of constraint applications
of ${\cal C}$.

\begin{enumerate}
\item If $U(x,y) \equiv \overline{x} \wedge y$, define
$\widehat{S}(\widehat{G})$ as 
$S(\widehat{G}) \cup \{U(f,t), U(f_1,t), U(f_2,t_1)\}$.
Then $\widehat{S}(\widehat{G})$ is equivalent to
$S(\widehat{G}) \cup \{\overline{f}, \overline{f_1}, \overline{f_2},
t, t_1\}$.

\item If $U(x,y) \equiv x \oplus y$, define
$\widehat{S}(\widehat{G})$ as 
$S(\widehat{G}) \cup \{(U(f,t), U(f_1,t), U(f_2,t), U(f,t_1)\}$.
Then $\widehat{S}(\widehat{G})$ is equivalent to
$S(\widehat{G}) \cup \{f \xor t, f_1 \xor t, f_2 \xor t,
f \xor t_1\}$.

\item If $U(x,y) \equiv x \leftrightarrow y$, define
$\widehat{S}(\widehat{G})$ as 
$S(\widehat{G}) \cup \{(U(f,f_1), U(f,f_2), U(t,t_1)\}$.
Then $\widehat{S}(\widehat{G})$ is equivalent to
$S(\widehat{G}) \cup \{f \leftrightarrow f_1, f \leftrightarrow f_2, t \leftrightarrow t_1\}$.

\item If $U(t,x,y) \equiv t \wedge (x \leftrightarrow y)$,
define $\widehat{S}(\widehat{G})$ as 
$S(\widehat{G}) \cup \{(U(t,f,f_1), U(t_1,f,f_2)\}$.
Then $\widehat{S}(\widehat{G})$ is equivalent to
$S(\widehat{G}) \cup \{t, t_1, (f \leftrightarrow f_1),
(f \leftrightarrow f_2)\}$.

\item If $U(f,x,y) \equiv \overline{f} \wedge (x \leftrightarrow y)$,
define $\widehat{S}(\widehat{G})$ as 
$S(\widehat{G}) \cup \{(U(f,f,f_1), U(f_2,t,t_1)\}$.
Then $\widehat{S}(\widehat{G})$ is equivalent to
$S(\widehat{G}) \cup \{\overline{f}, \overline{f_1}, \overline{f_2},
(t \leftrightarrow t_1)\}$.
\end{enumerate}

Let $G$ and $H$ be graphs such that
$V(G) = V(H) = \{1, \ldots, n\}$, $E(G) = \{e_1, \ldots, e_m\}$,
$E(H) = \{e'_1, \ldots, e'_m\}$,
all vertices in $G$ and $H$ have degree at least two,
and $G$ and $H$ do not contain triangles.
By Lemma~\ref{l:girestriction}, it suffices to show that
$G$ is isomorphic to $H$ if and only if
$\widehat{S}(G)$ is isomorphic to $\widehat{S}(H)$.

If there exists an isomorphism from $G$ to $H$, then,
by Observation~\ref{o:},
there  exists an isomorphism from  $S(G)$ to $S(H)$ that maps
$f$ to $f$ and $t$ to $t$. 
We can easily extend this to an isomorphism from 
$\widehat{S}(G)$ to $\widehat{S}(H)$, by mapping
$f_1$ to $f_1$, $f_2$ to $f_2$, and $t_1$ to $t_1$.

For the converse, note that, in all cases,
$\widehat{S}(\widehat{G}) \rightarrow
(f \leftrightarrow f_1) \wedge (f \leftrightarrow f_2) \wedge 
(t \leftrightarrow t_1)$.  Also note that in all cases,
$S(\widehat{G}) \cup \{\overline{f}, \overline{f_1}, \overline{f_2},
t, t_1\} \rightarrow \widehat{S}(\widehat{G})$.  

Now suppose that 
$\widehat{S}(G)$  is isomorphic to $\widehat{S}(H)$.
By Lemma~\ref{l:copies} and the observations made above,
for $\widehat{G} \in \{G,H\}$,
$\{f, f_1, f_2\}$ is the unique triple of equivalent
distinct variables. It follows that the isomorphism
maps $\{f, f_1, f_2\}$  to $\{f, f_1, f_2\}$.
It also follows from Lemma~\ref{l:copies} and the observations made
above that among the remaining variables, $\{t, t_1\}$ is the unique pair
of equivalent distinct variables. Thus, the isomorphism
maps $\{t, t_1\}$ to $\{t, t_1\}$.  It follows that 
$\widehat{S}(G)[f, f_1, f_2 := 0, t, t_1 := 1]$ is isomorphic to
$\widehat{S}(H)[f, f_1, f_2 := 0, t, t_1 := 1]$.

For $\widehat{G} \in \{G,H\}$,
$\widehat{S}(\widehat{G})[f, f_1, f_2 := 0, t, t_1 := 1]$ is equivalent
to $S(\widehat{G})[f := 0, t := 1]$. It follows that
$S(G)[f := 0, t := 1]$ is isomorphic to 
$S(H)[f := 0, t := 1]$. By Fact~\ref{f:gitoconst},
it follows that $G$ is isomorphic to $H$.
\end{prooftext}

\end{document}